\documentclass[aps,twocolumn,groupedaddress,superscriptaddress,floatfix,pra,longbibliography]{revtex4-1}
\usepackage[english]{babel} 
\usepackage{amsmath}
\usepackage{amssymb}
\usepackage{graphicx}
\usepackage{textcomp}
\usepackage{bm}	
\usepackage{soul}

\usepackage[usenames,dvipsnames]{color}	
\newcommand{\blu}[1]{\textcolor{black}{#1}}
\newcommand{\bblu}[1]{\textcolor{black}{#1}}
\newcommand{\red}[1]{\textcolor{black}{#1}}

\usepackage[braket, qm]{qcircuit}

\begin{document}

\title{Quantum implementation of an artificial feed-forward neural network}

\author{Francesco Tacchino}\email{francesco.tacchino01@ateneopv.it}
\affiliation{Dipartimento di Fisica, Universit\`a di Pavia, via Bassi 6, I-27100, Pavia, Italy}
\author{Panagiotis Barkoutsos}
\affiliation{IBM Research Zurich, S\"{a}umerstrasse 4, CH-8803 R\"{u}schlikon, Switzerland}
\author{Chiara Macchiavello}
\affiliation{Dipartimento di Fisica, Universit\`a di Pavia, via Bassi 6, I-27100, Pavia, Italy}
\affiliation{INFN Sezione di Pavia, via Bassi 6, I-27100, Pavia, Italy}
\affiliation{CNR-INO, largo E.\ Fermi 6, I-50125, Firenze, Italy }
\author{Ivano Tavernelli}
\affiliation{IBM Research Zurich, S\"{a}umerstrasse 4, CH-8803 R\"{u}schlikon, Switzerland}
\author{Dario Gerace}
\affiliation{Dipartimento di Fisica, Universit\`a di Pavia, via Bassi 6, I-27100, Pavia, Italy}
\author{Daniele Bajoni}
\affiliation{Dipartimento di Ingegneria Industriale e dell'Informazione, Universit\`a di Pavia, via Ferrata 1, I-27100, Pavia, Italy}

\pacs{xxxx, xxxx, xxxx}                                         
\date{\today}
\begin{abstract}

Artificial intelligence algorithms largely build on multi-layered neural networks. Coping with their increasing complexity and memory requirements calls for a paradigmatic change in the way these powerful algorithms are run. Quantum computing promises to solve certain tasks much more efficiently than any classical computing machine, and actual quantum processors are now becoming available through cloud access to perform experiments and testing also outside of research labs. \\
Here we show in practice an experimental realization of an artificial feed-forward neural network implemented on a state-of-art superconducting quantum processor using up to 7 active qubits. The network is made of quantum artificial neurons, which individually display a potential advantage in storage capacity with respect to their classical counterpart, and it is able to carry out an elementary classification task which would be impossible to achieve with a single node. We demonstrate that this network can be equivalently operated \bblu{either} via classical control or in a \bblu{completely} coherent fashion, thus opening the way to {hybrid} \bblu{as well as fully} quantum solutions for artificial intelligence to be run on near-term intermediate-scale quantum hardware.

\end{abstract}

\maketitle

\section{Introduction}

The field of artificial intelligence was revolutionized by moving from the simple, single layer perceptron design~\cite{Rosenblatt1957} to that of a complete feed-forward neural network (ffNN), constituted by several neurons organized in multiple successive layers~\cite{Hinton2006,Hinton2007}. In such artificial neural network designs each constituent neuron receives, as inputs, the outputs (activations) from the neurons in the preceding layer. The advantage of ffNNs with respect to simpler designs such as single layer perceptrons or support vector machines is that they can be used to classify data with relations that cannot be reduced to a separating hyperplane~\cite{Goodfellow-et-al-2016}. 
The present ubiquitous use of artificial intelligence in a wide variety of tasks, ranging from  pattern  or  spoken language recognition to the analysis of large data sets, is mostly due to the discovery that such feed-forward networks can be trained by using well established optimization algorithms~\cite{Hinton2006,Hinton2007,Goodfellow-et-al-2016}.

Quantum computers hold promise to achieve some form of computing advantage over classical counterparts in the not-so-far \red{future \cite{arute_quantum_2019}}. Indeed, quantum computing has been theoretically shown to offer potentially exponential speedups over traditional computing machines, especially in tasks such as large number factoring, solving linear systems of equations, and data classification~\bblu{\cite{NielsenChuang,shor_polynomial-time_1997,harrow_quantum_2009,lloyd_quantum_2014,rebentrost_quantum_2014}}. More recently, quantum computers have been applied to the field of Artificial Intelligence~\bblu{\cite{schuld_quest_2014,rebentrost_quantum_2018,Lloyd_quantum_algorithms_machine_learning_arxiv_2016,biamonte_quantum_2017}}, and recent realizations of artificial neurons~\cite{schuld_simulating_2015,schuld_implementing_2017,cao_quantum_2017,tacchino_artificial_2019} and support vector machines~\cite{Havlicek_Gambetta_qSVS_Nature_2019,schuld_quantum_2019} on real quantum processors, even if limited to simple systems at present, have shown a promising route towards a practical realization of such advantage.

In order to harness the full potentialities that quantum computing may offer to the field of artificial intelligence it is necessary to undergo the passage from single layered to deep feed-forward neural networks~\cite{wan_quantum_2017,grant_hierarchical_2018,cong_quantum_2019}, which has so greatly expanded the capabilities of artificially intelligent systems to date. Here we propose the architecture of a quantum ffNN and we test it on a state-of-the art 20-qubit IBMQ quantum processor. We start from a hybrid approach combining quantum nodes with classical information feed-forward, obtained via classical control of unitary transformations on qubits. This design realizes a fully general implementation of a ffNN on a quantum processor assisted by classical registers. A minimal 3-node example, specifically designed to carry out a pattern recognition task exceeding the capabilities of a single artificial neuron, is used for a proof-of-principle demonstration on real quantum hardware. We then describe and successfully implement on a 7-qubit register an equivalent fully quantum coherent configuration of the same set-up, which does not \bblu{involve} classical control of the feed-forward links and thus potentially opens the way to the exploration of more complex and classically inaccessible regimes.

\bblu{The proposed quantum implementation of ffNN offers interesting perspectives on scalability already in the Noisy Intermediate-Scale Quantum (NISQ)~\cite{preskill_quantum_2018} regime: indeed, the single quantum nodes potentially feature exponential advantage in memory usage, thus allowing to manipulate high-dimensional data structures with intermediate-size quantum registers, in principle. Moreover, the hybrid nature of the ffNN itself suggests a seamless integration with existing classical structures and algorithms for neural network computation and machine learning~\cite{mari_transfer_2019}.}

\section{Design of the hybrid feed-forward neural network}

In this section, we outline the general structure of our proposed hybrid ffNN, including a synthetic description of the working principles of single nodes and a more detailed discussion of layer-to-layer connections. While, for the sake of clarity, we will often refer to a specific minimal example with three nodes and two layers, the overall scheme can be generalized to arbitrary feed-forward networks.

\subsection{Individual nodes}

A ffNN is essentially composed of a set of individual nodes $\{n_i\} $, or artificial neurons, arranged in a set of successive layers $\{L_j\}$. Information flows through the network in a well defined direction from the input to the output layer, travelling through neuron-neuron connections (i.e.\ artificial synapses). Each node performs an elementary non-linear operation on the incoming data, whose result is then passed on to one or more nodes in the successive layer. 

In their simplest form, individual nodes can be designed to analyze binary-valued inputs. The artificial neurons that we consider here are based on the well known perceptron model~\cite{Rosenblatt1957}: such computational units analyze information by combining input ($\vec{i}$) and weight ($\vec{w}$) vectors, providing an activation response that depends on their scalar product $\vec{i}\cdot\vec{w}$. In our case, input and weight vectors are assumed to be binary-valued $m$-dimensional arrays\cite{McCulloch_Pitts_1943}, i.e.
\begin{equation}
\vec{i} = \begin{pmatrix}
    i_{0} \\
    i_{1} \\
    \vdots \\
    i_{m-1}
\end{pmatrix}\, ,\quad
\vec{w} = \begin{pmatrix}
    w_{0} \\
    w_{1} \\
    \vdots \\
    w_{m-1}
\end{pmatrix}
\end{equation}
where $i_k,w_k\in\{-1,1\}\,\forall k$. \bblu{The activity of a binary artificial neuron can be implemented on a quantum register of $N = \log_2(m)$ qubits~\cite{tacchino_artificial_2019}} by considering the quantum states
\begin{equation}
    |\psi_i\rangle = \frac{1}{\sqrt{m}}\sum_{j = 0}^{m - 1} i_j |j\rangle \, ,\quad
    |\psi_w\rangle = \frac{1}{\sqrt{m}}\sum_{j = 0}^{m - 1} w_j 
    |j\rangle
\label{eq:encodingstates}
\end{equation}
These encode the corresponding input and weight vectors by effectively exploiting the exponential size of the Hilbert space associated to the quantum register in use. The states of the form presented in Eq.~\eqref{eq:encodingstates} are real equally-weighted (REW) superpositions of all the computational basis states $|j\rangle \in \{|0\ldots 00\rangle ;|0\ldots 01\rangle ;\ldots,|1\ldots 11\rangle \}$. \bblu{The quantum procedure carrying out the perceptron-like computation for single artificial neurons can be summarized in three steps~\cite{tacchino_artificial_2019}.} First, assuming that the $N$-qubits quantum register is initially in the idle configuration, $|0\rangle^{\otimes N}$, we prepare the quantum state encoding the input vector with a unitary operation $\mathrm{U}_i$ such that $|\psi_i\rangle = \mathrm{U}_i|0\rangle^{\otimes N}$. \bblu{We then apply the weight factors of vector $\vec{w}$ on the input state by} implementing another unitary transformation, $\mathrm{U}_w$, subject to the constraint $|1\rangle^{\otimes N} = \mathrm{U}_w|\psi_w\rangle$. An optimized yet exact implementation of $\mathrm{U}_i$ and $\mathrm{U}_w$ exploits the close relationship between REW quantum states and the class of hypergraph states~\cite{Rossi2013,tacchino_artificial_2019}, achieving in the worst case an overall computational complexity which is linear in the size of the classical input, i.e.\ $O(m)$. After the two unitaries have been performed, it is easily seen that the state of the quantum register is
\begin{equation}
    |\phi_{i,w}\rangle = \mathrm{U}_w\mathrm{U}_i |0\rangle^{\otimes N} = \sum_{j = 0}^{m - 1} c_j |j\rangle
\end{equation}
where $c_{m-1} = \langle \psi_w | \psi_i\rangle = (1/m) \vec{i}\cdot\vec{w}$. Finally, the non-linear activation of the single artificial neuron can be implemented by performing a multi-controlled $\mathrm{NOT}$ gate~\cite{NielsenChuang} between the encoding register and an ancilla initialized in the initial state $|0\rangle$
\begin{equation}
    |\phi_{i,w}\rangle|0\rangle_a \rightarrow \sum_{j = 0}^{m - 2} c_j |j\rangle|0\rangle_a + c_{m-1}|m-1\rangle|1\rangle_a
    \label{eq:activation}
\end{equation}
followed by a final measure of the ancilla in the computational basis. Hence, the output of the quantum artificial neuron is found in the active state $|1\rangle_a$ with probability $p(1) = |c_{m-1}|^2$.

\subsection{Information feed-forward}

\begin{figure}
\begin{center}
\includegraphics[width=1\columnwidth]{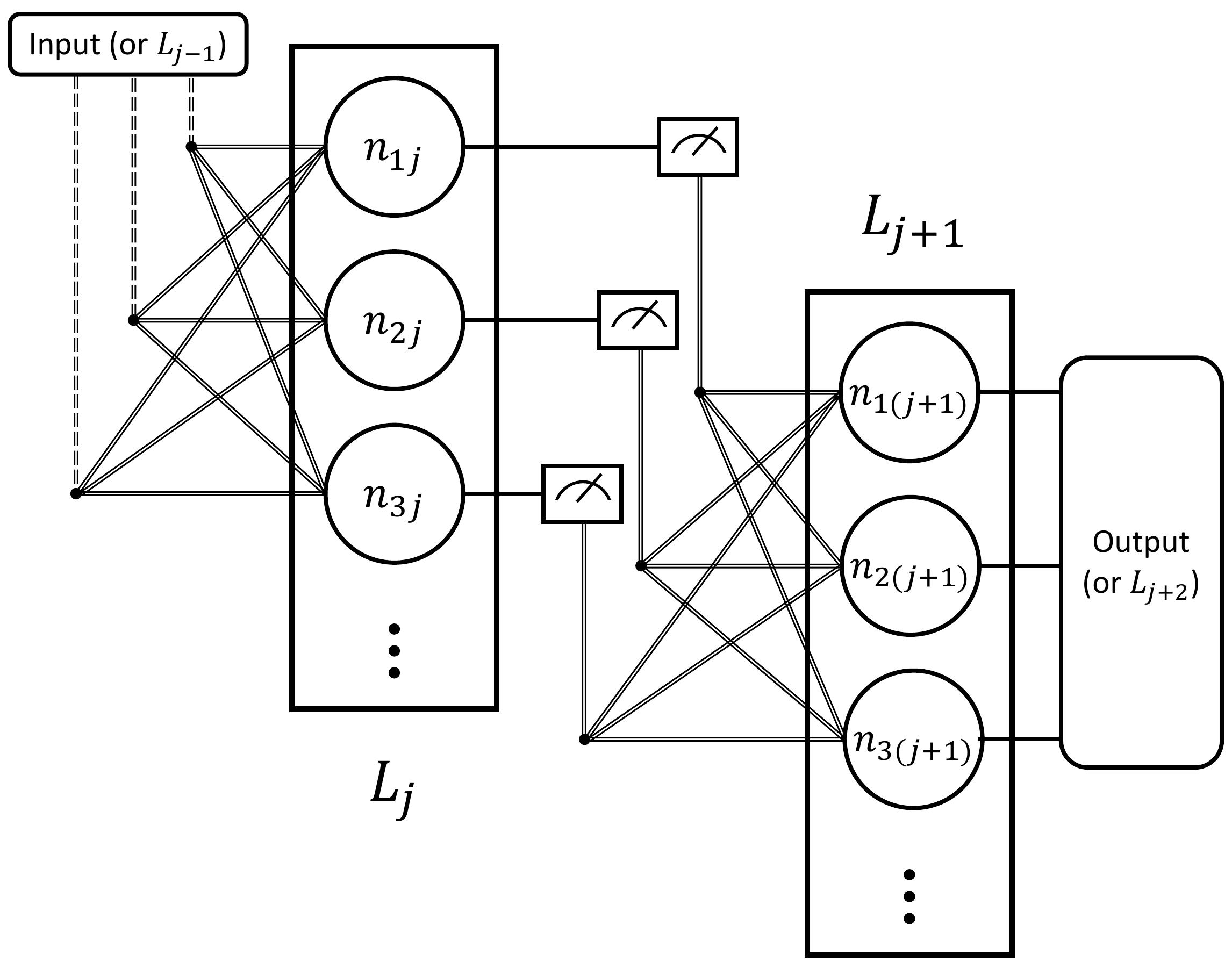} 
\caption{{\bf Abstract architecture of a hybrid ffNN.} Each layer $L_j$ contains an arbitrary number of nodes \bblu{$\{n_{kj}\}$}, which can individually be implemented on a quantum hardware. Upon measurement, information about the activation state of a layer is passed to the following one ($L_{j+1}$) in the form of classical bits controlling quantum operations. Full connectivity between nodes in successive layers is \red{schematically} shown, although sparser networks are also possible in principle. \bblu{The dashed line represents classical inputs from a generic preceding stage, which can be, e.g., a collection of layers up to $L_{j-1}$ or the original input information.}}
\label{fig:ffNN-abstract}
\end{center}
\end{figure}

When several copies of the quantum register implementing the artificial neuron model outlined above work in parallel, the respective ancillae, and the result of the measurements performed on them, can be used to feed-forward the information about the input-weight processing to a successive layer. Indeed, \red{let us} suppose that a layer $L_j$ contains \bblu{$\ell_j$} independent nodes, $\{n_{kj}\}_{k=1}^{\ell_j}$, each of them characterized by a weight vector $\vec{w}_{kj}$: in one cycle of operation, every node is provided with a classical input $\vec{i}_{kj}$ (either coming from layer $L_{j-1}$ or directly from the original data set to be analyzed) and, upon measurement, it outputs an activation state $a_{kj} \in \{1,0\}$, chosen according to a probability $p_{kj}(a_{kj}=1) \propto |\vec{i}_{kj}\cdot\vec{w}_{kj}|^2$. Assuming for simplicity that the $h$-th neuron $n_{h(j+1)}$ belonging to the $L_{j+1}$ layer collects the outputs of all $\{n_{kj}\}$ nodes, the corresponding binary classical input can be constructed as
\begin{equation}
    \vec{i}_{h(j+1)} = \begin{pmatrix}
    (-1)^{a_{1j}} \\
    (-1)^{a_{2j}} \\
    \vdots \\
    (-1)^{a_{\ell_jj}}
\end{pmatrix}
\end{equation}
Such new input vector can then be used to parametrize the appropriate $\mathrm{U}_i$ transformation for the $n_{h(j+1)}$ node. The overall computation can then be constructed by iteratively alternating the unitary quantum computation carried out by single layers with non-linear measurement and feed-forward stages. Notice that the design is totally general in terms of the number of nodes in each layer, the number of connections and the size of the various inputs to individual nodes. Moreover, as the information is formally transferred in the form of classical bits, the same input can easily be manipulated, e.g.,\ by making \bblu{classical} copies to be fed to independent nodes sharing similar connections to the previous layer. An abstract representation of the proposed architecture is shown in Fig.~\ref{fig:ffNN-abstract}.

From the technical point of view, a very natural implementation of the hybrid ffNN architecture onto a quantum processor makes use of classically controlled quantum gates. Independent quantum nodes within the same layer can either be implemented in different quantum registers, and thus computed simultaneously, or run on the same set of qubits, after proper re-initialization and by storing all the observed activation states in different positions of a classical memory register.

\subsection{Example: pattern recognition}
\label{sec:example_pattern_rec}

\begin{figure}
\begin{center}
\includegraphics[width=1\columnwidth]{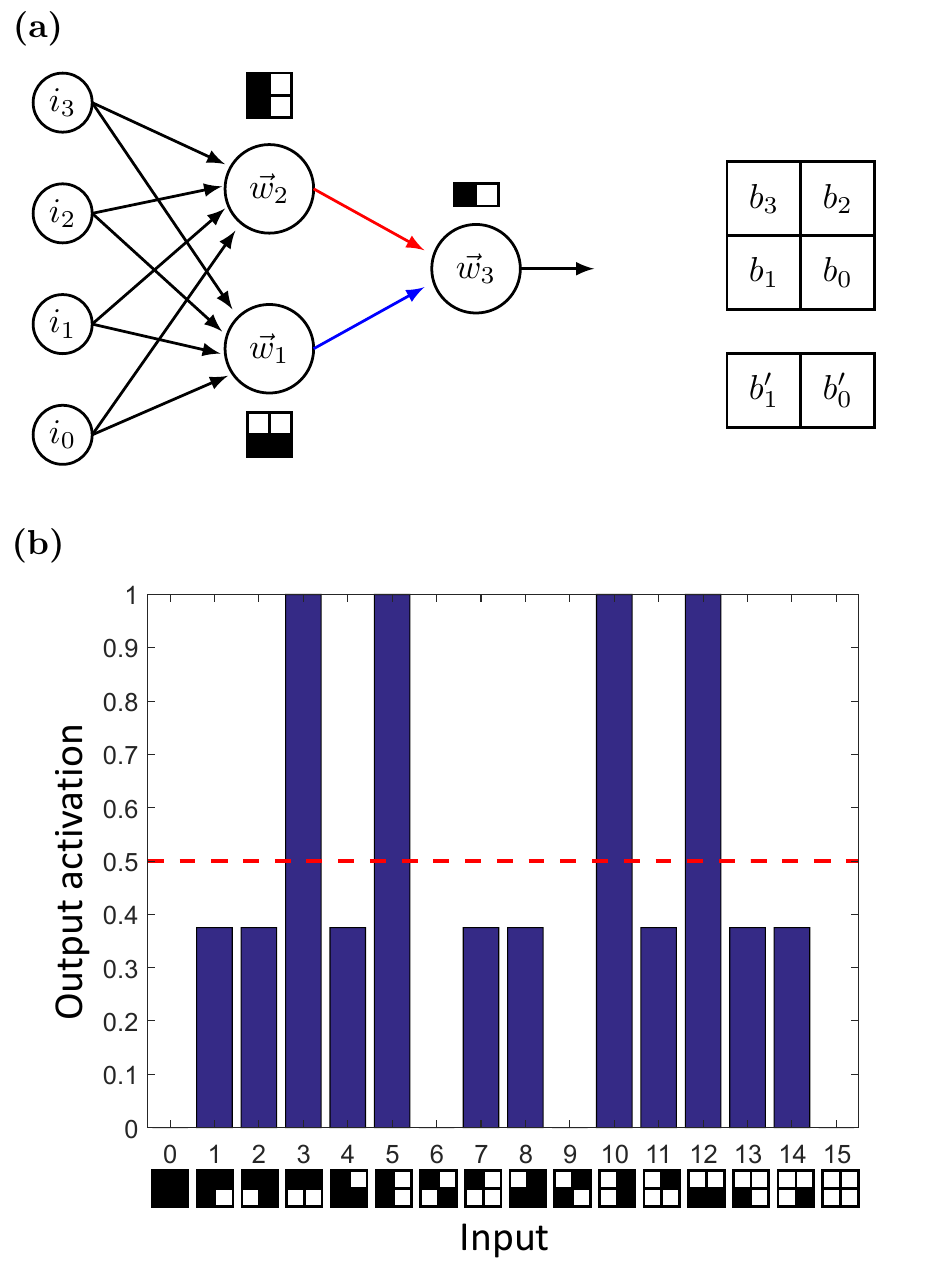} 
\caption{{\bf 3-node ffNN for patter recognition.} (a) The minimal example of a feed-forward neural network that we analyze in this study accepts four classical binary inputs and features one hidden layer containing two artificial neurons plus one output layer made of a single neuron. Next to each neuron, the ideal shape of the weight vectors achieving the desired recognition of horizontal and vertical lines is shown. The corresponding encoding scheme in terms of black and white pixels is also reported for a generic input/weight binary vector $\vec{b} = (b_0,\ldots,b_m)$. (b) Ideal results for the classification of $2\times 2$ pixel images. Notice that the target patterns, corresponding to integer labels $12$, $3$ (horizontal), $10$ and $5$ (vertical) all have $p^{out} = 1$, while all others have $p^{out} < 0.5$ (threshold shown in red). }
\label{fig:ffNN-3nodes-scheme}
\end{center}
\end{figure}

The working principles of our proposed hybrid ffNN, including the above technical details, are actually best clarified by describing an explicit example tailored to solve a well defined elementary classification problem. This will also set the stage for the experimental proof-of-principle demonstration on \red{an actual} superconducting \red{quantum} hardware to be presented in the next section. First, let us recall that binary input and weight vectors can be visually interpreted as images containing black or white square pixels~\cite{tacchino_artificial_2019}: a natural encoding scheme associates, e.g.,\ a white spot to a $i_j (w_j) = -1$ entry in the corresponding input (weight) vector, as shown explicitly in Fig.~\ref{fig:ffNN-3nodes-scheme}a \bblu{for the hidden ($m=4$, i.e.\ $2\times 2$ pixel images) and output ($m=2$, i.e.\ $2\times 1$ pixel images) layers of a minimal ffNN.} Moreover, we can identify any such binary pattern with a unique integer label by considering the equivalent decimal representation of the binary number $\mathtt{b}_3\mathtt{b}_2\mathtt{b}_1\mathtt{b}_0$ where $b_k = (-1)^{\mathtt{b}_k},\,\mathtt{b}_k\in\{0,1\}$. The task that we set out to solve with our example ffNN is the following: the network should be able to recognize (i.e., give a positive output activation with sufficiently large probability) whether there exist straight lines in $2\times 2$ pixel images, regardless of the fact that the lines are horizontal or vertical. All the other possible input images should be classified as negative. Notice that, as the data vectors encoding horizontal and vertical lines are orthogonal to each other, there is no single \blu{hyperplane} separating the four positive states from all other possible input images: therefore, the desired classification cannot be carried out by a single node accepting 4-bit inputs. This behavior of quantum artificial neurons differs from their usual classical counterparts, which cannot correctly classify sets containing opposite vectors~\cite{Goodfellow-et-al-2016}. More explicitly, given an input vector $\vec{v}_1$ and a weight vector $\vec{w}$, a single
quantum neuron would output a value proportional to $|\vec{v}_1\cdot\vec{w}|^2$, i.e.\ $\cos^2\theta$, where
$\theta$ is the angle formed by the two vectors. If we take a second input vector
$\vec{v}_2 \perp \vec{v}_1$, \bblu{the output would be upper bounded by $\sin^2\theta$.} As the set of patterns that should yield a positive result includes vectors that are orthogonal (those representing horizontal lines are orthogonal to those representing vertical lines) and vectors that are opposite (for instance, the vector corresponding to a vertical
line on the left column of a $2\times 2$ pixel image is opposite to the vector corresponding to a vertical line on the right column), it is therefore
impossible to find a weight $\vec{w}$ capable of yielding an output activation larger
than $0.5$ for all targets in the configuration space. \red{We hereby show that} a simple three-node network can accomplish the desired computation. A scheme of such \red{an elementary} ffNN is shown in Fig.~\ref{fig:ffNN-3nodes-scheme}a, where the circles indicate individual artificial neurons, and the vectors $\vec{w}_i$ refer to their respective weights. The network features a single hidden layer and a single binary output neuron. On a conceptual level, the functioning of the network can be interpreted as follows: with the a priori choice of weights represented in Fig.~\ref{fig:ffNN-3nodes-scheme}a, the top quantum neuron of the hidden layer outputs a high activation if the input vector has vertical lines, while the bottom neuron does the same for the case of horizontal lines. The output neuron in the last layer then recognizes whether one of the neurons in the hidden layer has given a positive outcome.

\begin{figure*}[t!]
\begin{center}
\includegraphics[width=\textwidth]{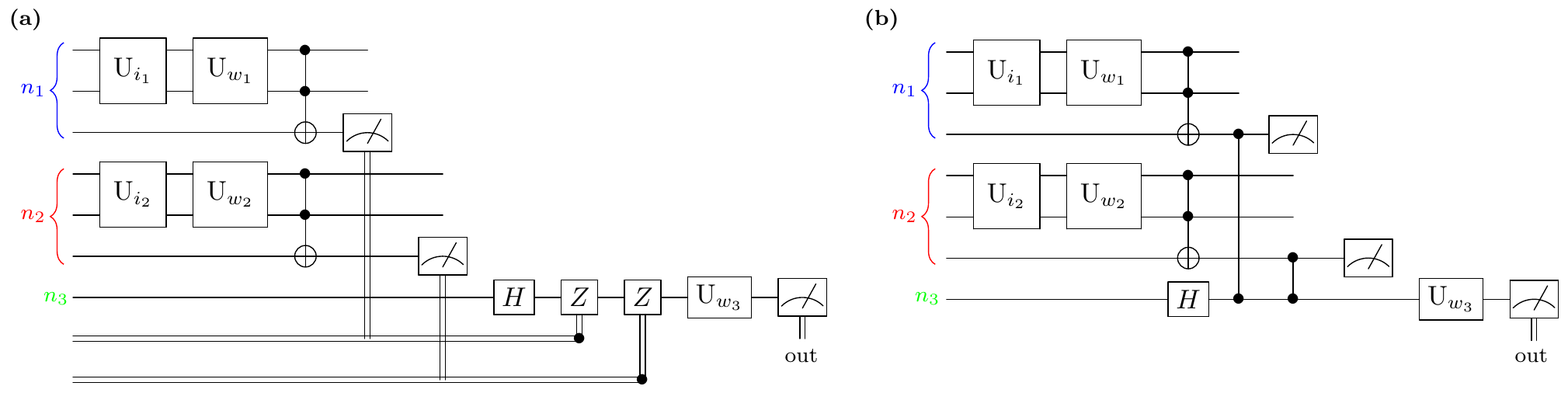} 
\caption{{\bf Circuit implementation of a ffNN.} (a) Hybrid realization of the feed-forward architecture introduced in Fig.~\ref{fig:ffNN-3nodes-scheme} via classical control. (b) Equivalent quantum coherent version using quantum controlled operations.}
\label{fig:ffNN-circuits}
\end{center}
\end{figure*}

A possible quantum circuit description of the ffNN introduced above, including the classical feed-forward stage between the hidden and the output layer, is provided in Fig.~\ref{fig:ffNN-circuits}a. We assume that each neuron within the hidden layer can accept \bblu{4-bit} inputs, such that each quantum neuron can be represented on a 2-qubit encoding register plus an ancilla qubit (i.e., $m=4$ and $N=2$ in this case). At the same time, the output neuron takes 2-dimensional inputs coming from the previous layer and provides the global activation state of the network, thus requiring a single qubit ($m=2$, $N=1$) to be encoded. Classical bits are also included to store the intermediate and final results.

Let us call $n_1$ and $n_2$ the two hidden nodes, which actually accept the same classical input but process it in two different ways. As described at the beginning of this section, each artificial neuron will independently provide, upon measurement, an activation pattern $a_k \in \{0,1\}\,(\text{for}\,k = 1,2)$, which can be stored in a classical bit $b_k$. We denote $p_k$ the probability of actually observing a value $a_k = 1$ from the $k$-th neuron. When such measurement is performed, we set $b_k = a_k$: as a result, the state of the classical 2-bit register after the quantum computation in the hidden layer has been completed is one of the following
\begin{equation}
    [b_1,b_2] = \begin{cases} 
    [0,0] \\
    [0,1] \\
    [1,0] \\
    [1,1]
    \end{cases}
\end{equation}
with probability
\begin{equation}
    p([b_1,b_2]) = \begin{cases} 
    (1-p_1)(1-p_2)\\
    (1-p_1)p_2\\
    p_1(1-p_2)\\
    p_1p_2
    \end{cases}
\label{eq:hiddenlayerprobs}
\end{equation}
respectively. It is easy to see that feed-forwarding the information contained in the classical register to the output neuron $n_3$ corresponds to providing it with one of the classical binary inputs $\vec{i}_{b_1b_2}$ reading
\begin{equation}
\begin{split}
    \vec{i}_{00}=\begin{pmatrix}
    1 \\
    1
    \end{pmatrix}, \quad
    \vec{i}_{01}=\begin{pmatrix}
    1 \\
    -1
    \end{pmatrix} \\
    \vec{i}_{10}=\begin{pmatrix}
    -1 \\
    1
    \end{pmatrix}, \quad
    \vec{i}_{11}=\begin{pmatrix}
    -1 \\
    -1
    \end{pmatrix}
\end{split}
\end{equation}

\noindent As shown in Fig.~\ref{fig:ffNN-circuits}a, a straightforward strategy for preparing the corresponding $|\psi_i\rangle$ state on the single-qubit register representing $n_3$ is by first bringing it from the idle state $|0\rangle$ to the superposition $\sqrt{2}|+\rangle = |0\rangle + |1\rangle$ via a Hadamard ($\mathrm{H}$) gate, and then conditioning the application of two $\mathrm{Z}$ gates (each of them adds a $-1$ phase to the $|1\rangle$ component, if applied) on the two classical bits $[b_1,b_2]$. The resulting quantum state will then be
\begin{equation}
    |\psi_i\rangle_{n_3} = \frac{1}{\sqrt{2}}\left(|0\rangle + (-1)^{b_1 \oplus b_2}|1\rangle\right)
\end{equation}
where $a\oplus b$ here denotes the usual bit sum modulo 2. If we now choose, as shown in Fig.~\ref{fig:ffNN-3nodes-scheme}a, a weight vector $\vec{w}_3 = (1,-1)$ we obtain $\mathrm{U}_{w_3} \equiv \mathrm{H}$. Therefore, the final state of the third neuron reads
\begin{equation}
    |\psi_{out}\rangle_{n_3} =
    \begin{cases}
    |0\rangle \quad & \text{if } b_1 \oplus b_2 = 0 \\
    |1\rangle \quad & \text{if } b_1 \oplus b_2 = 1
    \end{cases}
\label{eq:finalConditionedState}
\end{equation}
The overall probability of observing an active state on the output neuron can be written, in general, as
\begin{equation}
    p^{out} = \sum_{[b_1,b_2]} p([b_1,b_2])p(a_3 = 1 | [b_1,b_2])
\label{eq:class_trace_probs12}
\end{equation}
where we employed the usual notation for conditional probabilities and 
\begin{equation}
p(a_3 = 1 | [b_1,b_2]) = |\langle 1 | \psi_{out}\rangle_{n_3}|^2
\end{equation}
In our specific case, it is easy to see that, given Eq.~\eqref{eq:hiddenlayerprobs} and Eq.~\eqref{eq:finalConditionedState}, this reduces to
\begin{equation}
    p^{out} = p_1(1-p_2) + (1-p_1)p_2
\label{eq:ffnn_prob_convolution}
\end{equation}
Since in this elementary example $n_3$ is encoded in a single qubit, the final measurement can be performed directly without the need for an additional ancilla. In Fig.~\ref{fig:ffNN-3nodes-scheme}b we report the exact result for the convolution of Eq.~\eqref{eq:ffnn_prob_convolution}: as it can be seen, the ffNN ideally outputs an active state with $p^{out} = 1$ for the target horizontal and vertical patterns, while $p^{out} < 0.5$ in all other cases.

\red{Before moving forward, it is worth mentioning that the construction of a classically conditioned $\mathrm{U}_i$ can always be found also in more general cases, e.g.\ when the hidden layer contains more than two neurons. In particular, any node encoded on $N$ qubits will be able to accept inputs from $m = 2^N$ nodes in the previous layer: indeed, each output configuration from the latter will be one of the $2^m$ possible bit strings \bblu{$[b_1,\ldots,b_m]$} that can be used to uniquely identify one of the $2^m = 2^{2^N}$ possible input states, and thus to classically program its preparation.}

\bblu{\section{Quantum coherent feed-forward}}

The hybrid feed-forward architecture described so far and realized in a minimal 3-node 2-layer example can also be reformulated in a fully quantum coherent way. As we will show below, and at difference with the hybrid quantum-classical solution, this version always requires all nodes to be implemented simultaneously on a dedicated quantum register, thus making the quantum computation more demanding. At the same time, however, it reduces the necessity to store and process classical bits during intermediate stages. Moreover, fully coherent quantum neural networks offer more opportunities for use on quantum processors, as will be discussed in the final conclusions.

In Fig.~\ref{fig:ffNN-circuits}b we show a fully quantum construction for the ffNN of Fig.~\ref{fig:ffNN-circuits}a. The fundamental reason for the actual equivalence lies in the well known principle of deferred measurement~\cite{NielsenChuang}, stating that in a quantum circuit one can always move a measurement done at an intermediate stage to the end of the computation while replacing classically controlled operations ($\mathrm{O}$) with quantum controlled ones:
\begin{equation*}
    \Qcircuit @C=1.0em @R=0.0em @!R {
	 	& \meter & \control \cw \cwx[2] & & & & & & \ctrl{2} & \meter\\
	 	& & & & & = & & & & \\
	 	& \qw & \gate{\mathrm{O}} & \qw & & & & & \gate{\mathrm{O}} & \qw \\
	 }
\end{equation*}
Indeed, assuming that the nodes $n_1$ and $n_2$ are encoded in parallel and after the operations of the first layer (except the measurement on the ancillae) have been performed, we can write the global state of the total (3+3+1)-qubit network as
\begin{equation}
\begin{split}
        & \left(r_{n_1}|\varphi_{n_1}\rangle|0\rangle_{a_1} + c_{m-1,n_1}|1\ldots 1\rangle_{n_1}|1\rangle_{a_1}\right) \\
        & \otimes \left(r_{n_2}|\varphi_{n_2}\rangle|0\rangle_{a_2} + c_{m-1,n_2}|1\ldots 1\rangle_{n_2}|1\rangle_{a_2}\right) \\
        & \otimes |0\rangle_{n_3}
\end{split}
\end{equation}
where $r_{n_x}=(1-c_{m-1,n_x}^2)^{1/2}$ and $|\varphi_{n_x}\rangle$ contains, for each neuron, all the components other than the one leading to activation, see Eq.~\eqref{eq:activation}. Notice that, by construction, $\langle \varphi_{n_x} | 1\ldots1\rangle = 0$. In the meantime, the $n_3$ qubit is brought  into the superposition $\sqrt{2}|+\rangle = |0\rangle+|1\rangle$ by applying a single-qubit Hadamard gate, $\mathrm{H}$. Synapses can thereafter be implemented with two $\mathrm{CZ}$ gates, as represented in Fig.~\ref{fig:ffNN-circuits}b. The overall state of the quantum ffNN then becomes
\begin{equation}
\begin{split}
    & \left(r_{n_1}r_{n_2}|R_{n_1}\rangle|R_{n_2}\rangle + c_{n_1}c_{n_2}|A_{n_1}\rangle|A_{n_2}\rangle \right)|+\rangle_{n_3} \\
     & + \left(r_{n_1}c_{n_2}|R_{n_1}\rangle|A_{n_2}\rangle + c_{n_1}r_{n_2}|A_{n_1}\rangle|R_{n_2}\rangle \right)|-\rangle_{n_3}
     \label{eq:after_syn}
\end{split}
\end{equation}
where $c_{n_x}$ is a short-hand notation for $c_{m-1,n_x}$, and the activated $|A\rangle$ and rest $|R\rangle$ states of $n_1$ and $n_2$ are explicitly given as
\begin{equation}
\begin{split}
    |A_{n_x}\rangle & = |1\ldots 1\rangle_{n_x}|1\rangle_{a_x} \\
    |R_{n_x}\rangle & = |\varphi_{n_x}\rangle|0\rangle_{a_x}
\end{split}    
\end{equation}
By applying $\mathrm{U}_{w_3} \equiv \mathrm{H}$ on $n_3$ we obtain an output state
\begin{equation}
\begin{split}
    & |\psi_{out}\rangle = \\
    & \left(r_{n_1}r_{n_2}|R_{n_1}\rangle|R_{n_2}\rangle + c_{n_1}c_{n_2}|A_{n_1}\rangle|A_{n_2}\rangle \right)|0\rangle_{n_3} \\
     & + \left(r_{n_1}c_{n_2}|R_{n_1}\rangle|A_{n_2}\rangle + c_{n_1}r_{n_2}|A_{n_1}\rangle|R_{n_2}\rangle \right)|1\rangle_{n_3}
\end{split}
\end{equation}
It is straightforward to observe at this point that the neurons of the hidden layer can in principle be measured in an activation state $[b_1,b_2]\in\{[0,0],[0,1],[1,0],[1,1]\}$ with probabilities
\begin{equation}
    p([b_1,b_2]) = \begin{cases} 
    |r_{n_1}|^2|r_{n_2}|^2 = (1-p_1)(1-p_2)\\
    |r_{n_1}|^2|c_{n_2}|^2 = (1-p_1)p_2\\
    |c_{n_1}|^2|r_{n_2}|^2 = p_1(1-p_2)\\
    |c_{n_1}|^2|c_{n_2}|^2 = p_1p_2
    \end{cases}
\end{equation}
which exactly correspond to the ones reported in Eq.~\eqref{eq:hiddenlayerprobs}. However, as long as we are interested only in the output state of the network, i.e.\ the activation state $a_3$ of $n_3$, there is no need to actually perform the final measurements on $n_1$ and $n_2$: similarly to Eq.~\eqref{eq:class_trace_probs12}, we can in fact simply discard the information contained in the variables pertaining to the hidden layer by performing a partial trace operation. This returns a density matrix for the output neuron
\begin{equation}
    \rho_{n_3} = \operatorname{Tr}_{\{n_1,n_2\}}\left[|\psi_{out}\rangle \langle\psi_{out}|\right] = \begin{pmatrix}
    1-p^{out} & 0 \\
    0 & p^{out}
    \end{pmatrix}
\end{equation}
which automatically represents the convolution of the hidden nodes, see Eq.~\eqref{eq:ffnn_prob_convolution}. It is worth noticing that the role of the partial trace operation has recently been recognized and extensively discussed in the literature as a possible ingredient for a more general theory of quantum neural networks~\cite{torrontegui_unitary_2019,beer_efficient_2019}.

\bblu{To conclude this section, we also point out explicitly that the conversion between the two modes of operation (hybrid \textit{vs} coherent) of our proposed ffNN architecture goes beyond the specific example presented in this work. Indeed, as mentioned at the end of Sec.~\ref{sec:example_pattern_rec}, any feed-forward link between successive layers can in general be decomposed in terms of classically controlled operations. Whenever such construction is known, measurement deferral and partial traces can in principle always be employed to obtain the equivalent coherent network, namely by replacing all classical controls with their quantum counterparts and by measuring only the output layer.}
\begin{figure}
\begin{center}
\includegraphics[width=1\columnwidth]{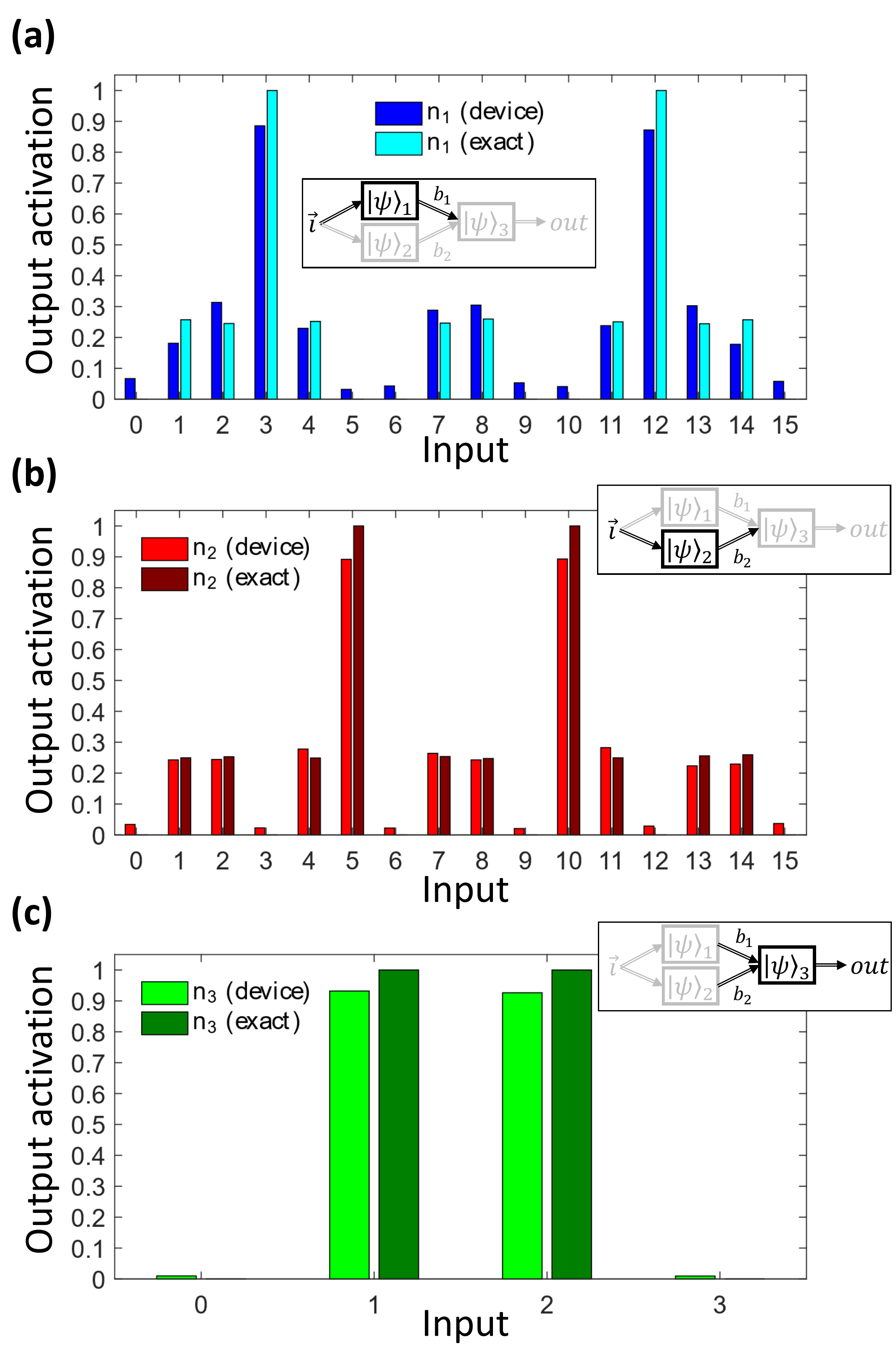} 
\caption{{\bf Experimental realization of single nodes on quantum hardware.} Single artificial neurons of the ffNN introduced in Fig.~\ref{fig:ffNN-circuits}a implemented on the IBMQ Poughkeepsie superconducting processor and compared with ideal noiseless outcomes computed numerically with the Qiskit \texttt{qasm\_simulator}. (a) Neuron $n_1$, recognizing horizontal inputs. (b) Neuron $n_2$, recognizing vertical inputs. (c) Neuron $n_3$, recognizing 2-dimensional inputs with dissimilar entries. Error mitigation is applied to data for $n_1$ and $n_2$.}
\label{fig:panel_single_neurons}
\end{center}
\end{figure}

\section{Experimental realization on a superconducting NISQ processor}

\begin{figure*}[t!]
\begin{center}
\includegraphics[width=\textwidth]{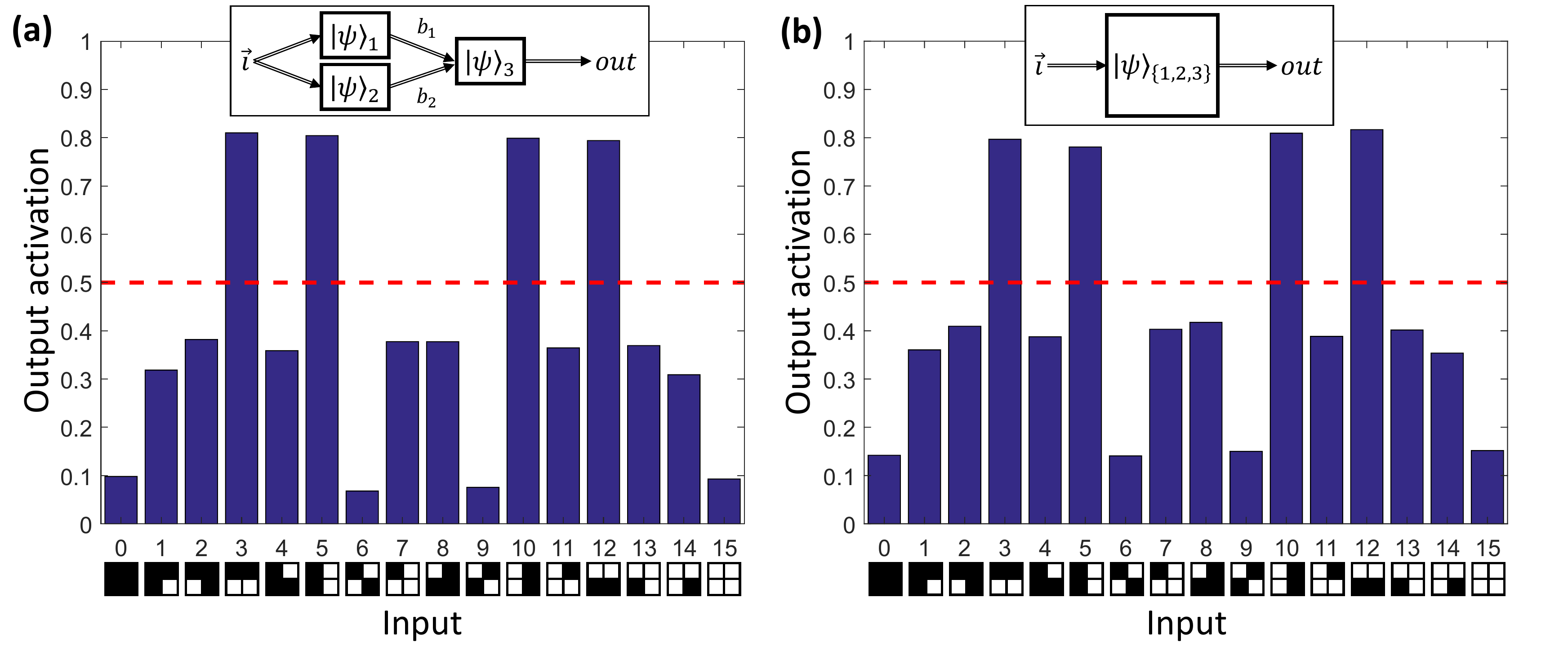} 
\caption{{\bf Results for the quantum ffNN classifying horizontal and vertical lines.} (a) Classification in the hybrid configuration, applying Eq.~\eqref{eq:class_trace_probs12} to the (error mitigated) experimental outcomes of Fig.~\ref{fig:panel_single_neurons}. (b) Classification in the coherent configuration obtained with a 7-qubit calculation on the IBMQ Poughkeepsie quantum processor (error mitigation is applied). Despite some residual quantitative inaccuracy, all the target patterns are correctly recognized if a threshold $\epsilon = 0.5$ (shown in red) is applied to the outcome probabilities both in the hybrid and the coherent versions. }
\label{fig:panel_full_exp}
\end{center}
\end{figure*}

We have implemented the ffNN introduced in Fig.~\ref{fig:ffNN-3nodes-scheme} on a real superconducting NISQ processor made available on cloud via the IBM Quantum Experience and programmed using the Qiskit python library~\cite{Qiskit}. Employing the same device, named IBMQ Poughkeepsie, we realized both the hybrid (Fig.~\ref{fig:ffNN-circuits}a) and the fully coherent (Fig.~\ref{fig:ffNN-circuits}b) configurations, reporting in both cases a remarkable successful completion of all the desired classification tasks.

In Fig.~\ref{fig:panel_single_neurons}a-b we show the results for the 3-qubit simulation of nodes $n_1$ and $n_2$,  respectively corresponding to the first and second set of three qubits in Fig.~\ref{fig:ffNN-circuits}a, from which the probabilities $p_1$ and $p_2$ can be estimated for all possible input vectors while assuming the weights $\vec{w}_1$ and $\vec{w}_2$ shown in Fig.~\ref{fig:ffNN-3nodes-scheme}a. The comparison with ideal results simulated numerically shows an excellent qualitative agreement and a  good quantitative \bblu{match} of the outcomes: in particular, notice that each individual node can successfully single out either vertical or horizontal lines\bblu{, see patterns in Fig.~\ref{fig:ffNN-3nodes-scheme}b}~\cite{tacchino_artificial_2019}. The agreement is naturally better for the simulation of all possible $n_3$ circuits, whose results are reported in Fig.~\ref{fig:panel_single_neurons}c: indeed, in this case the probability $p(a_3 = 1|[b_1,b_2])$ can be computed operating on a single qubit. The final outcomes (i.e.\ $p^{out}$) for the hybrid configuration of the ffNN, reported in Fig.~\ref{fig:panel_full_exp}a, are then obtained by applying Eq.~\eqref{eq:class_trace_probs12}. The latter is used in place of e.g.\ Eq.~\eqref{eq:ffnn_prob_convolution} in order to avoid introducing unnecessary assumptions or biases in the calculation and to take into account all possible sources of inaccuracy such as, for example, a non exactly zero outcome for $p(a_3 = 1|[0,0])$.

Finally, the experimental results for the fully coherent ffNN configuration are reported in Fig.~\ref{fig:panel_full_exp}b. These were obtained by running the 7-qubit quantum circuit introduced in Fig.~\ref{fig:ffNN-circuits}b. As it can immediately be appreciated, the outcomes are in good agreement with the corresponding ones in the hybrid version of the ffNN. We stress that such comparison is made non-trivial from the experimental point of view by the fact that, in the fully coherent version, a register of 7 simultaneously active and typically entangled qubits is required. On the contrary, the hybrid solution only requires each individual node to be separately implemented on a 3-qubit quantum register and, provided that the classical outcomes are conveniently stored, such quantum computations can be carried out in dedicated runs, thus avoiding e.g.\ cross-talks effects. As in the hybrid case, and despite some residual quantitative inaccuracy in the estimation of the activation probabilities, all the possible inputs are classified correctly by the ffNN, with the target horizontal and vertical patterns singled out from all other patterns. 

We also mention that raw data from the quantum processor already allow for an accurate classification in both hybrid and coherent configurations. However, the overall quality of the outcomes greatly benefits from the application of simple error mitigation techniques~\cite{temme_error_2017,li_efficient_2017,klco_quantum-classical_2018,kandala_error_2019}.

\section{Discussion}

In this work we have presented an original architecture to build feed-forward neural networks on universal quantum computing hardware \bblu{and demonstrated the use of them in NISQ devices}. In particular, we have shown how successive layers constituted by artificial neurons and implemented on independent quantum registers can be \red{either} connected to each other via classical control operations, thus realizing a hybrid quantum-classical ffNN, or by fully coherent quantum synapses. The necessary degree of non-linearity is achieved \red{in one case}  via explicit quantum measurement, \red{in the other} by a partial trace operation that effectively produces a convolution operation. \bblu{We stress that our proposed procedure is hardware-independent and therefore it can, in principle,} be implemented on any \red{quantum computing} machine, e.g. based on superconducting qubits~\cite{wendin_quantum_2017}, trapped-ions based quantum processors~\cite{schindler_quantum_2013}, and photonic components~\cite{AspuruGuzik:2012hoa,Takeda_review_CWPhotQC_APLPhot2019}.

\red{In the present work, we} have successfully tested a 3-node implementation of our algorithm applied to an elementary pattern classification task, both in the hybrid and fully coherent configurations. Such proof-of-principle demonstration was achieved on the IBMQ Poughkeepsie superconducting quantum processor by using up to 7 active qubits, and finding \red{a} substantial experimental agreement between the two proposed \red{operating modes} of the network. These results represent, to the best of our knowledge, one of the largest quantum neural network computation reported to date in terms of the total size of the quantum register. We also notice that the use of quantum artificial neurons as individual nodes gives the prospective advantage of an exponential gain in storage and processing ability: in turn, this confirms that hybrid quantum-classical neural networks could already be able to treat very large input vectors, beyond the capabilities of current systems. Such ability is becoming increasingly needed to handle e.g.\ very large image files, sanitary data for public health, market data for financial applications, and the ``data deluge'' expected from the Internet of Things. Moreover, the hybrid structure of our proposed ffNN could actually represent a relevant technical feature in the process of integrating quantum and classical processes for machine learning tasks: indeed, one could for example easily imagine that a few carefully distributed quantum nodes at the input of an otherwise classical network might act as a memory-efficient convolutional layer enabling the treatment of otherwise unmanageable sets of data.

A very natural extension of this work, and particularly of the fully coherent setup, would be an exploration of classically inaccessible regimes with no hybrid (i.e.\ classically controlled) counterpart. This could be achieved, e.g.,\ by allowing more complex synapse operations, thus letting activation probabilities for all neurons feeding the same successive layer to interfere in a truly quantum coherent way, or by engineering non-trivial quantum correlations between quantum nodes already within the same layer. In addition to the large advantage in data treatment capacity, this could then also result in new functionalities, such as the ability to deploy complicated convolution filters impossible to be run on classical hardware.

Even further reaching consequences might be expected from the possibility to directly process quantum data instead of quantum-encoded classical information, for instance to search for patterns in the output of a quantum simulator or process quantum states coming from a quantum internet appliance. In these cases, the input would directly be given in the form of a wavefunction or a density matrix~\cite{beer_efficient_2019}, without the resource cost associated to a classical input~\cite{Bergholm:2005:state_preparation_decomposition,Plesch:2011:state_preparation_decomposition_better,Mosca:01:state_preparation_multicontrolled}.

A last remark concerns the quantum network training. The practical example shown in this work used weights that were selected by the programmers, instead of discovering the weights through an optimization process (training). The nonlinearity coming from the measurement on the ancilla of each artificial neuron is sufficient,  in principle,  to \red{guarantee the required} plasticity for training~\cite{tacchino_artificial_2019}. This means that the architecture for quantum artificial neural networks proposed in this work is fully compatible with classical training algorithms, like the backpropagation method or the Newton-Raphson method~\cite{Goodfellow-et-al-2016}. However, one possible drawback of such methods is that they would incur in exponentially large training costs, i.e.\ when dealing with the very large vector spaces that could be associated to quantum neural networks. A possible alternative would be to use hybrid quantum-classical methods, like for instance Variational Quantum Eigensolvers~\cite{mcclean_theory_2016} to find the optimal weights. In particular, some VQE protocols have been shown to be implementable with an efficient (i.e. polynomial) use of classical resources~\cite{kandala_hardware-efficient_2017,moll_quantum_2018,kokail_self-verifying_2019} and some strategies have also been put forward to deal with the well known issue of barren plateaus in quantum neural networks~\cite{mcclean_barren_2018,grant_initialization_2019}. A thorough study of the training is however beyond the scope of the present work.

In conclusion, we provide a clear-cut recipe to map classical feed-forward neural networks onto quantum processors, and our results suggest that the whole design may eventually benefit from paradigmatic quantum properties such as superposition and entanglement. This represents a necessary step towards the final goal of approaching quantum advantage in the operation and training of quantum neural network applications.

\section{Acknowledgements}
We thank M.\ Fanizza and S.\ Woerner for useful discussions. We acknowledge the University of Pavia Blue Sky Research project number BSR1732907. This research was also supported by the Italian Ministry of Education, University and Research (MIUR): ``Dipartimenti di Eccellenza Program (2018-2022)'', Department of Physics, University of Pavia and PRIN Project INPhoPOL.


\begin{thebibliography}{47}%
\makeatletter
\providecommand \@ifxundefined [1]{%
 \@ifx{#1\undefined}
}%
\providecommand \@ifnum [1]{%
 \ifnum #1\expandafter \@firstoftwo
 \else \expandafter \@secondoftwo
 \fi
}%
\providecommand \@ifx [1]{%
 \ifx #1\expandafter \@firstoftwo
 \else \expandafter \@secondoftwo
 \fi
}%
\providecommand \natexlab [1]{#1}%
\providecommand \enquote  [1]{``#1''}%
\providecommand \bibnamefont  [1]{#1}%
\providecommand \bibfnamefont [1]{#1}%
\providecommand \citenamefont [1]{#1}%
\providecommand \href@noop [0]{\@secondoftwo}%
\providecommand \href [0]{\begingroup \@sanitize@url \@href}%
\providecommand \@href[1]{\@@startlink{#1}\@@href}%
\providecommand \@@href[1]{\endgroup#1\@@endlink}%
\providecommand \@sanitize@url [0]{\catcode `\\12\catcode `\$12\catcode
  `\&12\catcode `\#12\catcode `\^12\catcode `\_12\catcode `\%12\relax}%
\providecommand \@@startlink[1]{}%
\providecommand \@@endlink[0]{}%
\providecommand \url  [0]{\begingroup\@sanitize@url \@url }%
\providecommand \@url [1]{\endgroup\@href {#1}{\urlprefix }}%
\providecommand \urlprefix  [0]{URL }%
\providecommand \Eprint [0]{\href }%
\providecommand \doibase [0]{http://dx.doi.org/}%
\providecommand \selectlanguage [0]{\@gobble}%
\providecommand \bibinfo  [0]{\@secondoftwo}%
\providecommand \bibfield  [0]{\@secondoftwo}%
\providecommand \translation [1]{[#1]}%
\providecommand \BibitemOpen [0]{}%
\providecommand \bibitemStop [0]{}%
\providecommand \bibitemNoStop [0]{.\EOS\space}%
\providecommand \EOS [0]{\spacefactor3000\relax}%
\providecommand \BibitemShut  [1]{\csname bibitem#1\endcsname}%
\let\auto@bib@innerbib\@empty
\bibitem [{\citenamefont {Rosenblatt}(1957)}]{Rosenblatt1957}%
  \BibitemOpen
  \bibfield  {author} {\bibinfo {author} {\bibfnamefont {F.}~\bibnamefont
  {Rosenblatt}},\ }\href@noop {} {\emph {\bibinfo {title} {The Perceptron: A
  perceiving and recognizing automaton}}},\ \bibinfo {type} {Tech. Rep.}\
  \bibinfo {number} {Inc. Report No. 85-460-1}\ (\bibinfo  {institution}
  {Cornell Aeronautical Laboratory},\ \bibinfo {year} {1957})\BibitemShut
  {NoStop}%
\bibitem [{\citenamefont {Hinton}\ \emph {et~al.}(2006)\citenamefont {Hinton},
  \citenamefont {Osindero},\ and\ \citenamefont {Teh}}]{Hinton2006}%
  \BibitemOpen
  \bibfield  {author} {\bibinfo {author} {\bibfnamefont {G.~E.}\ \bibnamefont
  {Hinton}}, \bibinfo {author} {\bibfnamefont {S.}~\bibnamefont {Osindero}}, \
  and\ \bibinfo {author} {\bibfnamefont {Y.-W.}\ \bibnamefont {Teh}},\
  }\bibfield  {title} {\enquote {\bibinfo {title} {A fast learning algorithm
  for deep belief nets},}\ }\href@noop {} {\bibfield  {journal} {\bibinfo
  {journal} {Neural Computation}\ }\textbf {\bibinfo {volume} {18}},\ \bibinfo
  {pages} {1527--1554} (\bibinfo {year} {2006})}\BibitemShut {NoStop}%
\bibitem [{\citenamefont {Hinton}(2007)}]{Hinton2007}%
  \BibitemOpen
  \bibfield  {author} {\bibinfo {author} {\bibfnamefont {G.~E.}\ \bibnamefont
  {Hinton}},\ }\bibfield  {title} {\enquote {\bibinfo {title} {Learning
  multiple layers of representation},}\ }\href@noop {} {\bibfield  {journal}
  {\bibinfo  {journal} {Trends in Cognitive Sciences}\ }\textbf {\bibinfo
  {volume} {11}},\ \bibinfo {pages} {428 -- 434} (\bibinfo {year}
  {2007})}\BibitemShut {NoStop}%
\bibitem [{\citenamefont {Goodfellow}\ \emph {et~al.}(2016)\citenamefont
  {Goodfellow}, \citenamefont {Bengio},\ and\ \citenamefont
  {Courville}}]{Goodfellow-et-al-2016}%
  \BibitemOpen
  \bibfield  {author} {\bibinfo {author} {\bibfnamefont {I.}~\bibnamefont
  {Goodfellow}}, \bibinfo {author} {\bibfnamefont {Y.}~\bibnamefont {Bengio}},
  \ and\ \bibinfo {author} {\bibfnamefont {A.}~\bibnamefont {Courville}},\
  }\href@noop {} {\emph {\bibinfo {title} {Deep Learning}}}\ (\bibinfo
  {publisher} {MIT Press},\ \bibinfo {year} {2016})\BibitemShut {NoStop}%
\bibitem [{\citenamefont {Arute~et al.}(2019)}]{arute_quantum_2019}%
  \BibitemOpen
  \bibfield  {author} {\bibinfo {author} {\bibfnamefont {F.}~\bibnamefont
  {Arute~et al.}},\ }\bibfield  {title} {\enquote {\bibinfo {title} {Quantum
  supremacy using a programmable superconducting processor},}\ }\href@noop {}
  {\bibfield  {journal} {\bibinfo  {journal} {Nature}\ }\textbf {\bibinfo
  {volume} {574}},\ \bibinfo {pages} {505--510} (\bibinfo {year}
  {2019})}\BibitemShut {NoStop}%
\bibitem [{\citenamefont {Nielsen}\ and\ \citenamefont
  {Chuang}(2004)}]{NielsenChuang}%
  \BibitemOpen
  \bibfield  {author} {\bibinfo {author} {\bibfnamefont {M.~A.}\ \bibnamefont
  {Nielsen}}\ and\ \bibinfo {author} {\bibfnamefont {I.~L.}\ \bibnamefont
  {Chuang}},\ }\href@noop {} {\emph {\bibinfo {title} {Quantum Computation and
  Quantum Information (Cambridge Series on Information and the Natural
  Sciences)}}},\ \bibinfo {edition} {1st}\ ed.\ (\bibinfo  {publisher}
  {Cambridge University Press},\ \bibinfo {year} {2004})\BibitemShut {NoStop}%
\bibitem [{\citenamefont {Shor}(1997)}]{shor_polynomial-time_1997}%
  \BibitemOpen
  \bibfield  {author} {\bibinfo {author} {\bibfnamefont {P.~W.}\ \bibnamefont
  {Shor}},\ }\bibfield  {title} {\enquote {\bibinfo {title} {Polynomial-{Time}
  {Algorithms} for {Prime} {Factorization} and {Discrete} {Logarithms} on a
  {Quantum} {Computer}},}\ }\href@noop {} {\bibfield  {journal} {\bibinfo
  {journal} {SIAM Journal on Computing}\ }\textbf {\bibinfo {volume} {26}},\
  \bibinfo {pages} {1484--1509} (\bibinfo {year} {1997})}\BibitemShut {NoStop}%
\bibitem [{\citenamefont {Harrow}\ \emph {et~al.}(2009)\citenamefont {Harrow},
  \citenamefont {Hassidim},\ and\ \citenamefont {Lloyd}}]{harrow_quantum_2009}%
  \BibitemOpen
  \bibfield  {author} {\bibinfo {author} {\bibfnamefont {A.~W.}\ \bibnamefont
  {Harrow}}, \bibinfo {author} {\bibfnamefont {A.}~\bibnamefont {Hassidim}}, \
  and\ \bibinfo {author} {\bibfnamefont {S.}~\bibnamefont {Lloyd}},\ }\bibfield
   {title} {\enquote {\bibinfo {title} {Quantum {Algorithm} for {Linear}
  {Systems} of {Equations}},}\ }\href@noop {} {\bibfield  {journal} {\bibinfo
  {journal} {Physical Review Letters}\ }\textbf {\bibinfo {volume} {103}},\
  \bibinfo {pages} {150502} (\bibinfo {year} {2009})}\BibitemShut {NoStop}%
\bibitem [{\citenamefont {Lloyd}\ \emph {et~al.}(2014)\citenamefont {Lloyd},
  \citenamefont {Mohseni},\ and\ \citenamefont
  {Rebentrost}}]{lloyd_quantum_2014}%
  \BibitemOpen
  \bibfield  {author} {\bibinfo {author} {\bibfnamefont {S.}~\bibnamefont
  {Lloyd}}, \bibinfo {author} {\bibfnamefont {M.}~\bibnamefont {Mohseni}}, \
  and\ \bibinfo {author} {\bibfnamefont {P.}~\bibnamefont {Rebentrost}},\
  }\bibfield  {title} {\enquote {\bibinfo {title} {Quantum principal component
  analysis},}\ }\href@noop {} {\bibfield  {journal} {\bibinfo  {journal}
  {Nature Physics}\ }\textbf {\bibinfo {volume} {10}},\ \bibinfo {pages}
  {631--633} (\bibinfo {year} {2014})}\BibitemShut {NoStop}%
\bibitem [{\citenamefont {Rebentrost}\ \emph {et~al.}(2014)\citenamefont
  {Rebentrost}, \citenamefont {Mohseni},\ and\ \citenamefont
  {Lloyd}}]{rebentrost_quantum_2014}%
  \BibitemOpen
  \bibfield  {author} {\bibinfo {author} {\bibfnamefont {P.}~\bibnamefont
  {Rebentrost}}, \bibinfo {author} {\bibfnamefont {M.}~\bibnamefont {Mohseni}},
  \ and\ \bibinfo {author} {\bibfnamefont {S.}~\bibnamefont {Lloyd}},\
  }\bibfield  {title} {\enquote {\bibinfo {title} {Quantum {Support} {Vector}
  {Machine} for {Big} {Data} {Classification}},}\ }\href@noop {} {\bibfield
  {journal} {\bibinfo  {journal} {Physical Review Letters}\ }\textbf {\bibinfo
  {volume} {113}},\ \bibinfo {pages} {130503} (\bibinfo {year}
  {2014})}\BibitemShut {NoStop}%
\bibitem [{\citenamefont {Schuld}\ \emph {et~al.}(2014)\citenamefont {Schuld},
  \citenamefont {Sinayskiy},\ and\ \citenamefont
  {Petruccione}}]{schuld_quest_2014}%
  \BibitemOpen
  \bibfield  {author} {\bibinfo {author} {\bibfnamefont {M.}~\bibnamefont
  {Schuld}}, \bibinfo {author} {\bibfnamefont {I.}~\bibnamefont {Sinayskiy}}, \
  and\ \bibinfo {author} {\bibfnamefont {F.}~\bibnamefont {Petruccione}},\
  }\bibfield  {title} {\enquote {\bibinfo {title} {The quest for a {Quantum}
  {Neural} {Network}},}\ }\href@noop {} {\bibfield  {journal} {\bibinfo
  {journal} {Quantum Information Processing}\ }\textbf {\bibinfo {volume}
  {13}},\ \bibinfo {pages} {2567--2586} (\bibinfo {year} {2014})}\BibitemShut
  {NoStop}%
\bibitem [{\citenamefont {Rebentrost}\ \emph {et~al.}(2018)\citenamefont
  {Rebentrost}, \citenamefont {Bromley}, \citenamefont {Weedbrook},\ and\
  \citenamefont {Lloyd}}]{rebentrost_quantum_2018}%
  \BibitemOpen
  \bibfield  {author} {\bibinfo {author} {\bibfnamefont {P.}~\bibnamefont
  {Rebentrost}}, \bibinfo {author} {\bibfnamefont {T.~R.}\ \bibnamefont
  {Bromley}}, \bibinfo {author} {\bibfnamefont {C.}~\bibnamefont {Weedbrook}},
  \ and\ \bibinfo {author} {\bibfnamefont {S.}~\bibnamefont {Lloyd}},\
  }\bibfield  {title} {\enquote {\bibinfo {title} {Quantum {Hopfield} neural
  network},}\ }\href@noop {} {\bibfield  {journal} {\bibinfo  {journal}
  {Physical Review A}\ }\textbf {\bibinfo {volume} {98}},\ \bibinfo {pages}
  {042308} (\bibinfo {year} {2018})}\BibitemShut {NoStop}%
\bibitem [{\citenamefont {Lloyd}\ \emph {et~al.}(2013)\citenamefont {Lloyd},
  \citenamefont {Mohseni},\ and\ \citenamefont
  {Rebentrost}}]{Lloyd_quantum_algorithms_machine_learning_arxiv_2016}%
  \BibitemOpen
  \bibfield  {author} {\bibinfo {author} {\bibfnamefont {S.}~\bibnamefont
  {Lloyd}}, \bibinfo {author} {\bibfnamefont {M.}~\bibnamefont {Mohseni}}, \
  and\ \bibinfo {author} {\bibfnamefont {P.}~\bibnamefont {Rebentrost}},\
  }\href@noop {} {\enquote {\bibinfo {title} {Quantum algorithms for supervised
  and unsupervised machine learning},}\ } (\bibinfo {year} {2013}),\ \Eprint
  {} {arXiv:1307.0411} \BibitemShut
  {NoStop}%
\bibitem [{\citenamefont {Biamonte}\ \emph {et~al.}(2017)\citenamefont
  {Biamonte}, \citenamefont {Wittek}, \citenamefont {Pancotti}, \citenamefont
  {Rebentrost}, \citenamefont {Wiebe},\ and\ \citenamefont
  {Lloyd}}]{biamonte_quantum_2017}%
  \BibitemOpen
  \bibfield  {author} {\bibinfo {author} {\bibfnamefont {J.}~\bibnamefont
  {Biamonte}}, \bibinfo {author} {\bibfnamefont {P.}~\bibnamefont {Wittek}},
  \bibinfo {author} {\bibfnamefont {N.}~\bibnamefont {Pancotti}}, \bibinfo
  {author} {\bibfnamefont {P.}~\bibnamefont {Rebentrost}}, \bibinfo {author}
  {\bibfnamefont {N.}~\bibnamefont {Wiebe}}, \ and\ \bibinfo {author}
  {\bibfnamefont {S.}~\bibnamefont {Lloyd}},\ }\bibfield  {title} {\enquote
  {\bibinfo {title} {Quantum machine learning},}\ }\href@noop {} {\bibfield
  {journal} {\bibinfo  {journal} {Nature}\ }\textbf {\bibinfo {volume} {549}},\
  \bibinfo {pages} {195} (\bibinfo {year} {2017})}\BibitemShut {NoStop}%
\bibitem [{\citenamefont {Schuld}\ \emph {et~al.}(2015)\citenamefont {Schuld},
  \citenamefont {Sinayskiy},\ and\ \citenamefont
  {Petruccione}}]{schuld_simulating_2015}%
  \BibitemOpen
  \bibfield  {author} {\bibinfo {author} {\bibfnamefont {M.}~\bibnamefont
  {Schuld}}, \bibinfo {author} {\bibfnamefont {I.}~\bibnamefont {Sinayskiy}}, \
  and\ \bibinfo {author} {\bibfnamefont {F.}~\bibnamefont {Petruccione}},\
  }\bibfield  {title} {\enquote {\bibinfo {title} {Simulating a perceptron on a
  quantum computer},}\ }\href@noop {} {\bibfield  {journal} {\bibinfo
  {journal} {Physics Letters A}\ }\textbf {\bibinfo {volume} {379}},\ \bibinfo
  {pages} {660--663} (\bibinfo {year} {2015})}\BibitemShut {NoStop}%
\bibitem [{\citenamefont {Schuld}\ \emph {et~al.}(2017)\citenamefont {Schuld},
  \citenamefont {Fingerhuth},\ and\ \citenamefont
  {Petruccione}}]{schuld_implementing_2017}%
  \BibitemOpen
  \bibfield  {author} {\bibinfo {author} {\bibfnamefont {M.}~\bibnamefont
  {Schuld}}, \bibinfo {author} {\bibfnamefont {M.}~\bibnamefont {Fingerhuth}},
  \ and\ \bibinfo {author} {\bibfnamefont {F.}~\bibnamefont {Petruccione}},\
  }\bibfield  {title} {\enquote {\bibinfo {title} {Implementing a
  distance-based classifier with a quantum interference circuit},}\ }\href@noop
  {} {\bibfield  {journal} {\bibinfo  {journal} {EPL (Europhysics Letters)}\
  }\textbf {\bibinfo {volume} {119}},\ \bibinfo {pages} {60002} (\bibinfo
  {year} {2017})}\BibitemShut {NoStop}%
\bibitem [{\citenamefont {Cao}\ \emph {et~al.}(2017)\citenamefont {Cao},
  \citenamefont {Guerreschi},\ and\ \citenamefont
  {Aspuru-Guzik}}]{cao_quantum_2017}%
  \BibitemOpen
  \bibfield  {author} {\bibinfo {author} {\bibfnamefont {Y.}~\bibnamefont
  {Cao}}, \bibinfo {author} {\bibfnamefont {G.~G.}\ \bibnamefont {Guerreschi}},
  \ and\ \bibinfo {author} {\bibfnamefont {A.}~\bibnamefont {Aspuru-Guzik}},\
  }\bibfield  {title} {\enquote {\bibinfo {title} {Quantum {Neuron}: an
  elementary building block for machine learning on quantum computers},}\
  }\href@noop {} {\bibfield  {journal} {\bibinfo  {journal} {arXiv:1711.11240
  [quant-ph]}\ } (\bibinfo {year} {2017})}\BibitemShut {NoStop}%
\bibitem [{\citenamefont {Tacchino}\ \emph {et~al.}(2019)\citenamefont
  {Tacchino}, \citenamefont {Macchiavello}, \citenamefont {Gerace},\ and\
  \citenamefont {Bajoni}}]{tacchino_artificial_2019}%
  \BibitemOpen
  \bibfield  {author} {\bibinfo {author} {\bibfnamefont {F.}~\bibnamefont
  {Tacchino}}, \bibinfo {author} {\bibfnamefont {C.}~\bibnamefont
  {Macchiavello}}, \bibinfo {author} {\bibfnamefont {D.}~\bibnamefont
  {Gerace}}, \ and\ \bibinfo {author} {\bibfnamefont {D.}~\bibnamefont
  {Bajoni}},\ }\bibfield  {title} {\enquote {\bibinfo {title} {An artificial
  neuron implemented on an actual quantum processor},}\ }\href@noop {}
  {\bibfield  {journal} {\bibinfo  {journal} {npj Quantum Information}\
  }\textbf {\bibinfo {volume} {5}},\ \bibinfo {pages} {26} (\bibinfo {year}
  {2019})}\BibitemShut {NoStop}%
\bibitem [{\citenamefont {Havl{\'\i}{\v c}ek}\ \emph
  {et~al.}(2019)\citenamefont {Havl{\'\i}{\v c}ek}, \citenamefont
  {C{\'o}rcoles}, \citenamefont {Temme}, \citenamefont {Harrow}, \citenamefont
  {Kandala}, \citenamefont {Chow},\ and\ \citenamefont
  {Gambetta}}]{Havlicek_Gambetta_qSVS_Nature_2019}%
  \BibitemOpen
  \bibfield  {author} {\bibinfo {author} {\bibfnamefont {V.}~\bibnamefont
  {Havl{\'\i}{\v c}ek}}, \bibinfo {author} {\bibfnamefont {A.~D.}\ \bibnamefont
  {C{\'o}rcoles}}, \bibinfo {author} {\bibfnamefont {K.}~\bibnamefont {Temme}},
  \bibinfo {author} {\bibfnamefont {A.~W.}\ \bibnamefont {Harrow}}, \bibinfo
  {author} {\bibfnamefont {A.}~\bibnamefont {Kandala}}, \bibinfo {author}
  {\bibfnamefont {J.~M.}\ \bibnamefont {Chow}}, \ and\ \bibinfo {author}
  {\bibfnamefont {J.~M.}\ \bibnamefont {Gambetta}},\ }\bibfield  {title}
  {\enquote {\bibinfo {title} {Supervised learning with quantum-enhanced
  feature spaces},}\ }\href@noop {} {\bibfield  {journal} {\bibinfo  {journal}
  {Nature}\ }\textbf {\bibinfo {volume} {567}},\ \bibinfo {pages} {209--212}
  (\bibinfo {year} {2019})}\BibitemShut {NoStop}%
\bibitem [{\citenamefont {Schuld}\ and\ \citenamefont
  {Killoran}(2019)}]{schuld_quantum_2019}%
  \BibitemOpen
  \bibfield  {author} {\bibinfo {author} {\bibfnamefont {M.}~\bibnamefont
  {Schuld}}\ and\ \bibinfo {author} {\bibfnamefont {N.}~\bibnamefont
  {Killoran}},\ }\bibfield  {title} {\enquote {\bibinfo {title} {Quantum
  {Machine} {Learning} in {Feature} {Hilbert} {Spaces}},}\ }\href@noop {}
  {\bibfield  {journal} {\bibinfo  {journal} {Physical Review Letters}\
  }\textbf {\bibinfo {volume} {122}},\ \bibinfo {pages} {040504} (\bibinfo
  {year} {2019})}\BibitemShut {NoStop}%
\bibitem [{\citenamefont {Wan}\ \emph {et~al.}(2017)\citenamefont {Wan},
  \citenamefont {Dahlsten}, \citenamefont {Kristj{\'a}nsson}, \citenamefont
  {Gardner},\ and\ \citenamefont {Kim}}]{wan_quantum_2017}%
  \BibitemOpen
  \bibfield  {author} {\bibinfo {author} {\bibfnamefont {K.~H.}\ \bibnamefont
  {Wan}}, \bibinfo {author} {\bibfnamefont {O.}~\bibnamefont {Dahlsten}},
  \bibinfo {author} {\bibfnamefont {H.}~\bibnamefont {Kristj{\'a}nsson}},
  \bibinfo {author} {\bibfnamefont {R.}~\bibnamefont {Gardner}}, \ and\
  \bibinfo {author} {\bibfnamefont {M.~S.}\ \bibnamefont {Kim}},\ }\bibfield
  {title} {\enquote {\bibinfo {title} {Quantum generalisation of feedforward
  neural networks},}\ }\href@noop {} {\bibfield  {journal} {\bibinfo  {journal}
  {npj Quantum Information}\ }\textbf {\bibinfo {volume} {3}},\ \bibinfo
  {pages} {36} (\bibinfo {year} {2017})}\BibitemShut {NoStop}%
\bibitem [{\citenamefont {Grant}\ \emph {et~al.}(2018)\citenamefont {Grant},
  \citenamefont {Benedetti}, \citenamefont {Cao}, \citenamefont {Hallam},
  \citenamefont {Lockhart}, \citenamefont {Stojevic}, \citenamefont {Green},\
  and\ \citenamefont {Severini}}]{grant_hierarchical_2018}%
  \BibitemOpen
  \bibfield  {author} {\bibinfo {author} {\bibfnamefont {E.}~\bibnamefont
  {Grant}}, \bibinfo {author} {\bibfnamefont {M.}~\bibnamefont {Benedetti}},
  \bibinfo {author} {\bibfnamefont {S.}~\bibnamefont {Cao}}, \bibinfo {author}
  {\bibfnamefont {A.}~\bibnamefont {Hallam}}, \bibinfo {author} {\bibfnamefont
  {J.}~\bibnamefont {Lockhart}}, \bibinfo {author} {\bibfnamefont
  {V.}~\bibnamefont {Stojevic}}, \bibinfo {author} {\bibfnamefont {A.~G.}\
  \bibnamefont {Green}}, \ and\ \bibinfo {author} {\bibfnamefont
  {S.}~\bibnamefont {Severini}},\ }\bibfield  {title} {\enquote {\bibinfo
  {title} {Hierarchical quantum classifiers},}\ } {\bibfield  {journal} {\bibinfo  {journal} {npj
  Quantum Information}\ }\textbf {\bibinfo {volume} {4}},\ \bibinfo {pages}
  {65} (\bibinfo {year} {2018})}\BibitemShut {NoStop}%
\bibitem [{\citenamefont {Cong}\ \emph {et~al.}(2019)\citenamefont {Cong},
  \citenamefont {Choi},\ and\ \citenamefont {Lukin}}]{cong_quantum_2019}%
  \BibitemOpen
  \bibfield  {author} {\bibinfo {author} {\bibfnamefont {I.}~\bibnamefont
  {Cong}}, \bibinfo {author} {\bibfnamefont {S.}~\bibnamefont {Choi}}, \ and\
  \bibinfo {author} {\bibfnamefont {M.~D.}\ \bibnamefont {Lukin}},\ }\bibfield
  {title} {\enquote {\bibinfo {title} {Quantum convolutional neural
  networks},}\ }\href@noop {} {\bibfield  {journal} {\bibinfo  {journal}
  {Nature Physics}\ }\textbf {\bibinfo {volume} {15}} (\bibinfo {year}
  {2019})}\BibitemShut {NoStop}%
\bibitem [{\citenamefont {Preskill}(2018)}]{preskill_quantum_2018}%
  \BibitemOpen
  \bibfield  {author} {\bibinfo {author} {\bibfnamefont {J.}~\bibnamefont
  {Preskill}},\ }\bibfield  {title} {\enquote {\bibinfo {title} {Quantum
  {Computing} in the {NISQ} era and beyond},}\ }\href@noop {} {\bibfield
  {journal} {\bibinfo  {journal} {Quantum}\ }\textbf {\bibinfo {volume} {2}}
  (\bibinfo {year} {2018})}\BibitemShut {NoStop}%
\bibitem [{\citenamefont {Mari}\ \emph {et~al.}(2019)\citenamefont {Mari},
  \citenamefont {Bromley}, \citenamefont {Izaac}, \citenamefont {Schuld},\ and\
  \citenamefont {Killoran}}]{mari_transfer_2019}%
  \BibitemOpen
  \bibfield  {author} {\bibinfo {author} {\bibfnamefont {A.}~\bibnamefont
  {Mari}}, \bibinfo {author} {\bibfnamefont {T.~R.}\ \bibnamefont {Bromley}},
  \bibinfo {author} {\bibfnamefont {J.}~\bibnamefont {Izaac}}, \bibinfo
  {author} {\bibfnamefont {M.}~\bibnamefont {Schuld}}, \ and\ \bibinfo {author}
  {\bibfnamefont {N.}~\bibnamefont {Killoran}},\ }\bibfield  {title} {\enquote
  {\bibinfo {title} {Transfer learning in hybrid classical-quantum neural
  networks},}\ }\href@noop {} {\bibfield  {journal} {\bibinfo  {journal}
  {arXiv:1912.08278 [quant-ph, stat]}\ } (\bibinfo {year} {2019})}\BibitemShut
  {NoStop}%
\bibitem [{\citenamefont {McCulloch}\ and\ \citenamefont
  {Pitts}(1943)}]{McCulloch_Pitts_1943}%
  \BibitemOpen
  \bibfield  {author} {\bibinfo {author} {\bibfnamefont {W.~S.}\ \bibnamefont
  {McCulloch}}\ and\ \bibinfo {author} {\bibfnamefont {W.}~\bibnamefont
  {Pitts}},\ }\bibfield  {title} {\enquote {\bibinfo {title} {A logical
  calculus of the ideas immanent in nervous activity},}\ }\href@noop {}
  {\bibfield  {journal} {\bibinfo  {journal} {The bulletin of mathematical
  biophysics}\ }\textbf {\bibinfo {volume} {5}},\ \bibinfo {pages} {115--133}
  (\bibinfo {year} {1943})}\BibitemShut {NoStop}%
\bibitem [{\citenamefont {Rossi}\ \emph {et~al.}(2013)\citenamefont {Rossi},
  \citenamefont {Huber}, \citenamefont {Bru{\ss}},\ and\ \citenamefont
  {Macchiavello}}]{Rossi2013}%
  \BibitemOpen
  \bibfield  {author} {\bibinfo {author} {\bibfnamefont {M.}~\bibnamefont
  {Rossi}}, \bibinfo {author} {\bibfnamefont {M.}~\bibnamefont {Huber}},
  \bibinfo {author} {\bibfnamefont {D.}~\bibnamefont {Bru{\ss}}}, \ and\
  \bibinfo {author} {\bibfnamefont {C.}~\bibnamefont {Macchiavello}},\
  }\bibfield  {title} {\enquote {\bibinfo {title} {Quantum hypergraph
  states},}\ }\href@noop {} {\bibfield  {journal} {\bibinfo  {journal} {New
  Journal of Physics}\ }\textbf {\bibinfo {volume} {15}},\ \bibinfo {pages}
  {113022} (\bibinfo {year} {2013})}\BibitemShut {NoStop}%
\bibitem [{\citenamefont {Torrontegui}\ and\ \citenamefont
  {Garcia-Ripoll}(2019)}]{torrontegui_unitary_2019}%
  \BibitemOpen
  \bibfield  {author} {\bibinfo {author} {\bibfnamefont {E.}~\bibnamefont
  {Torrontegui}}\ and\ \bibinfo {author} {\bibfnamefont {J.~J.}\ \bibnamefont
  {Garcia-Ripoll}},\ }\bibfield  {title} {\enquote {\bibinfo {title} {Unitary
  quantum perceptron as efficient universal approximator},}\ }\href@noop {}
  {\bibfield  {journal} {\bibinfo  {journal} {EPL (Europhysics Letters)}\
  }\textbf {\bibinfo {volume} {125}} (\bibinfo {year} {2019})}\BibitemShut
  {NoStop}%
\bibitem [{\citenamefont {Beer}\ \emph {et~al.}(2019)\citenamefont {Beer},
  \citenamefont {Bondarenko}, \citenamefont {Farrelly}, \citenamefont
  {Osborne}, \citenamefont {Salzmann},\ and\ \citenamefont
  {Wolf}}]{beer_efficient_2019}%
  \BibitemOpen
  \bibfield  {author} {\bibinfo {author} {\bibfnamefont {K.}~\bibnamefont
  {Beer}}, \bibinfo {author} {\bibfnamefont {D.}~\bibnamefont {Bondarenko}},
  \bibinfo {author} {\bibfnamefont {T.}~\bibnamefont {Farrelly}}, \bibinfo
  {author} {\bibfnamefont {T.~J.}\ \bibnamefont {Osborne}}, \bibinfo {author}
  {\bibfnamefont {R.}~\bibnamefont {Salzmann}}, \ and\ \bibinfo {author}
  {\bibfnamefont {R.}~\bibnamefont {Wolf}},\ }\bibfield  {title} {\enquote
  {\bibinfo {title} {Efficient {Learning} for {Deep} {Quantum} {Neural}
  {Networks}},}\ }\href@noop {} {\bibfield  {journal} {\bibinfo  {journal}
  {arXiv:1902.10445 [physics, physics:quant-ph]}\ } (\bibinfo {year}
  {2019})}\BibitemShut {NoStop}%
\bibitem [{\citenamefont {et~al.}(2019)}]{Qiskit}%
  \BibitemOpen
  \bibfield  {author} {\bibinfo {author} {\bibfnamefont {G.~Aleksandrowicz}\
  \bibnamefont {et~al.}},\ }\href {\doibase 10.5281/zenodo.2562110} {\enquote
  {\bibinfo {title} {Qiskit: An open-source framework for quantum computing},}\
  } (\bibinfo {year} {2019})\BibitemShut {NoStop}%
\bibitem [{\citenamefont {Temme}\ \emph {et~al.}(2017)\citenamefont {Temme},
  \citenamefont {Bravyi},\ and\ \citenamefont {Gambetta}}]{temme_error_2017}%
  \BibitemOpen
  \bibfield  {author} {\bibinfo {author} {\bibfnamefont {K.}~\bibnamefont
  {Temme}}, \bibinfo {author} {\bibfnamefont {S.}~\bibnamefont {Bravyi}}, \
  and\ \bibinfo {author} {\bibfnamefont {J.~M.}\ \bibnamefont {Gambetta}},\
  }\bibfield  {title} {\enquote {\bibinfo {title} {Error {Mitigation} for
  {Short}-{Depth} {Quantum} {Circuits}},}\ }\href@noop {} {\bibfield  {journal}
  {\bibinfo  {journal} {Physical Review Letters}\ }\textbf {\bibinfo {volume}
  {119}},\ \bibinfo {pages} {180509} (\bibinfo {year} {2017})}\BibitemShut
  {NoStop}%
\bibitem [{\citenamefont {Li}\ and\ \citenamefont
  {Benjamin}(2017)}]{li_efficient_2017}%
  \BibitemOpen
  \bibfield  {author} {\bibinfo {author} {\bibfnamefont {Y.}~\bibnamefont
  {Li}}\ and\ \bibinfo {author} {\bibfnamefont {S.~C.}\ \bibnamefont
  {Benjamin}},\ }\bibfield  {title} {\enquote {\bibinfo {title} {Efficient
  {Variational} {Quantum} {Simulator} {Incorporating} {Active} {Error}
  {Minimization}},}\ }\href@noop {} {\bibfield  {journal} {\bibinfo  {journal}
  {Physical Review X}\ }\textbf {\bibinfo {volume} {7}},\ \bibinfo {pages}
  {021050} (\bibinfo {year} {2017})}\BibitemShut {NoStop}%
\bibitem [{\citenamefont {Klco}\ \emph {et~al.}(2018)\citenamefont {Klco},
  \citenamefont {Dumitrescu}, \citenamefont {McCaskey}, \citenamefont {Morris},
  \citenamefont {Pooser}, \citenamefont {Sanz}, \citenamefont {Solano},
  \citenamefont {Lougovski},\ and\ \citenamefont
  {Savage}}]{klco_quantum-classical_2018}%
  \BibitemOpen
  \bibfield  {author} {\bibinfo {author} {\bibfnamefont {N.}~\bibnamefont
  {Klco}}, \bibinfo {author} {\bibfnamefont {E.~F.}\ \bibnamefont
  {Dumitrescu}}, \bibinfo {author} {\bibfnamefont {A.~J.}\ \bibnamefont
  {McCaskey}}, \bibinfo {author} {\bibfnamefont {T.~D.}\ \bibnamefont
  {Morris}}, \bibinfo {author} {\bibfnamefont {R.~C.}\ \bibnamefont {Pooser}},
  \bibinfo {author} {\bibfnamefont {M.}~\bibnamefont {Sanz}}, \bibinfo {author}
  {\bibfnamefont {E.}~\bibnamefont {Solano}}, \bibinfo {author} {\bibfnamefont
  {P.}~\bibnamefont {Lougovski}}, \ and\ \bibinfo {author} {\bibfnamefont
  {M.~J.}\ \bibnamefont {Savage}},\ }\bibfield  {title} {\enquote {\bibinfo
  {title} {Quantum-classical computation of {Schwinger} model dynamics using
  quantum computers},}\ }\href@noop {} {\bibfield  {journal} {\bibinfo
  {journal} {Physical Review A}\ }\textbf {\bibinfo {volume} {98}},\ \bibinfo
  {pages} {032331} (\bibinfo {year} {2018})}\BibitemShut {NoStop}%
\bibitem [{\citenamefont {Kandala}\ \emph {et~al.}(2019)\citenamefont
  {Kandala}, \citenamefont {Temme}, \citenamefont {C{\'o}rcoles}, \citenamefont
  {Mezzacapo}, \citenamefont {Chow},\ and\ \citenamefont
  {Gambetta}}]{kandala_error_2019}%
  \BibitemOpen
  \bibfield  {author} {\bibinfo {author} {\bibfnamefont {A.}~\bibnamefont
  {Kandala}}, \bibinfo {author} {\bibfnamefont {K.}~\bibnamefont {Temme}},
  \bibinfo {author} {\bibfnamefont {A.~D.}\ \bibnamefont {C{\'o}rcoles}},
  \bibinfo {author} {\bibfnamefont {A.}~\bibnamefont {Mezzacapo}}, \bibinfo
  {author} {\bibfnamefont {J.~M.}\ \bibnamefont {Chow}}, \ and\ \bibinfo
  {author} {\bibfnamefont {J.~M.}\ \bibnamefont {Gambetta}},\ }\bibfield
  {title} {\enquote {\bibinfo {title} {Error mitigation extends the
  computational reach of a noisy quantum processor},}\ }\href@noop {}
  {\bibfield  {journal} {\bibinfo  {journal} {Nature}\ }\textbf {\bibinfo
  {volume} {567}},\ \bibinfo {pages} {491--495} (\bibinfo {year}
  {2019})}\BibitemShut {NoStop}%
\bibitem [{\citenamefont {Wendin}(2017)}]{wendin_quantum_2017}%
  \BibitemOpen
  \bibfield  {author} {\bibinfo {author} {\bibfnamefont {G.}~\bibnamefont
  {Wendin}},\ }\bibfield  {title} {\enquote {\bibinfo {title} {Quantum
  information processing with superconducting circuits: a review},}\
  }\href@noop {} {\bibfield  {journal} {\bibinfo  {journal} {Rep. Prog. Phys.}\
  }\textbf {\bibinfo {volume} {80}},\ \bibinfo {pages} {106001} (\bibinfo
  {year} {2017})}\BibitemShut {NoStop}%
\bibitem [{\citenamefont {Schindler}\ \emph {et~al.}(2013)\citenamefont
  {Schindler}, \citenamefont {Nigg}, \citenamefont {Monz}, \citenamefont
  {Barreiro}, \citenamefont {Martinez}, \citenamefont {Wang}, \citenamefont
  {Quint}, \citenamefont {Brandl}, \citenamefont {Nebendahl}, \citenamefont
  {Roos}, \citenamefont {Chwalla}, \citenamefont {Hennrich},\ and\
  \citenamefont {Blatt}}]{schindler_quantum_2013}%
  \BibitemOpen
  \bibfield  {author} {\bibinfo {author} {\bibfnamefont {P.}~\bibnamefont
  {Schindler}}, \bibinfo {author} {\bibfnamefont {D.}~\bibnamefont {Nigg}},
  \bibinfo {author} {\bibfnamefont {T.}~\bibnamefont {Monz}}, \bibinfo {author}
  {\bibfnamefont {J.~T.}\ \bibnamefont {Barreiro}}, \bibinfo {author}
  {\bibfnamefont {E.}~\bibnamefont {Martinez}}, \bibinfo {author}
  {\bibfnamefont {S.~X.}\ \bibnamefont {Wang}}, \bibinfo {author}
  {\bibfnamefont {S.}~\bibnamefont {Quint}}, \bibinfo {author} {\bibfnamefont
  {M.~F.}\ \bibnamefont {Brandl}}, \bibinfo {author} {\bibfnamefont
  {V.}~\bibnamefont {Nebendahl}}, \bibinfo {author} {\bibfnamefont {C.~F.}\
  \bibnamefont {Roos}}, \bibinfo {author} {\bibfnamefont {M.}~\bibnamefont
  {Chwalla}}, \bibinfo {author} {\bibfnamefont {M.}~\bibnamefont {Hennrich}}, \
  and\ \bibinfo {author} {\bibfnamefont {R.}~\bibnamefont {Blatt}},\ }\bibfield
   {title} {\enquote {\bibinfo {title} {A quantum information processor with
  trapped ions},}\ }\href@noop {} {\bibfield  {journal} {\bibinfo  {journal}
  {New J. Phys.}\ }\textbf {\bibinfo {volume} {15}},\ \bibinfo {pages} {123012}
  (\bibinfo {year} {2013})}\BibitemShut {NoStop}%
\bibitem [{\citenamefont {Aspuru-Guzik}\ and\ \citenamefont
  {Walther}(2012)}]{AspuruGuzik:2012hoa}%
  \BibitemOpen
  \bibfield  {author} {\bibinfo {author} {\bibfnamefont {A.}~\bibnamefont
  {Aspuru-Guzik}}\ and\ \bibinfo {author} {\bibfnamefont {P.}~\bibnamefont
  {Walther}},\ }\bibfield  {title} {\enquote {\bibinfo {title} {{Photonic
  quantum simulators}},}\ }\href@noop {} {\bibfield  {journal} {\bibinfo
  {journal} {Nature Physics}\ }\textbf {\bibinfo {volume} {8}},\ \bibinfo
  {pages} {285--291} (\bibinfo {year} {2012})}\BibitemShut {NoStop}%
\bibitem [{\citenamefont {Takeda}\ and\ \citenamefont
  {Furusawa}(2019)}]{Takeda_review_CWPhotQC_APLPhot2019}%
  \BibitemOpen
  \bibfield  {author} {\bibinfo {author} {\bibfnamefont {S.}~\bibnamefont
  {Takeda}}\ and\ \bibinfo {author} {\bibfnamefont {A.}~\bibnamefont
  {Furusawa}},\ }\bibfield  {title} {\enquote {\bibinfo {title} {Toward
  large-scale fault-tolerant universal photonic quantum computing},}\ } {\bibfield  {journal} {\bibinfo  {journal} {APL
  Photonics}\ }\textbf {\bibinfo {volume} {4}},\ \bibinfo {pages} {060902}
  (\bibinfo {year} {2019})}\BibitemShut {NoStop}%
\bibitem [{\citenamefont {Bergholm}\ \emph {et~al.}(2005)\citenamefont
  {Bergholm}, \citenamefont {Vartiainen}, \citenamefont {M\"ott\"onen},\ and\
  \citenamefont {Salomaa}}]{Bergholm:2005:state_preparation_decomposition}%
  \BibitemOpen
  \bibfield  {author} {\bibinfo {author} {\bibfnamefont {V.}~\bibnamefont
  {Bergholm}}, \bibinfo {author} {\bibfnamefont {J.~J.}\ \bibnamefont
  {Vartiainen}}, \bibinfo {author} {\bibfnamefont {M.}~\bibnamefont
  {M\"ott\"onen}}, \ and\ \bibinfo {author} {\bibfnamefont {M.~M.}\
  \bibnamefont {Salomaa}},\ }\bibfield  {title} {\enquote {\bibinfo {title}
  {Quantum circuits with uniformly controlled one-qubit gates},}\ }\href@noop
  {} {\bibfield  {journal} {\bibinfo  {journal} {Phys. Rev. A}\ }\textbf
  {\bibinfo {volume} {71}},\ \bibinfo {pages} {052330} (\bibinfo {year}
  {2005})}\BibitemShut {NoStop}%
\bibitem [{\citenamefont {Plesch}\ and\ \citenamefont
  {Brukner}(2011)}]{Plesch:2011:state_preparation_decomposition_better}%
  \BibitemOpen
  \bibfield  {author} {\bibinfo {author} {\bibfnamefont {M.}~\bibnamefont
  {Plesch}}\ and\ \bibinfo {author} {\bibfnamefont {C.}~\bibnamefont
  {Brukner}},\ }\bibfield  {title} {\enquote {\bibinfo {title} {Quantum-state
  preparation with universal gate decompositions},}\ }\href@noop {} {\bibfield
  {journal} {\bibinfo  {journal} {Phys. Rev. A}\ }\textbf {\bibinfo {volume}
  {83}},\ \bibinfo {pages} {032302} (\bibinfo {year} {2011})}\BibitemShut
  {NoStop}%
\bibitem [{\citenamefont {Mosca}\ and\ \citenamefont
  {Kaye}(2001)}]{Mosca:01:state_preparation_multicontrolled}%
  \BibitemOpen
  \bibfield  {author} {\bibinfo {author} {\bibfnamefont {M.}~\bibnamefont
  {Mosca}}\ and\ \bibinfo {author} {\bibfnamefont {P.}~\bibnamefont {Kaye}},\
  }\bibfield  {title} {\enquote {\bibinfo {title} {Quantum networks for
  generating arbitrary quantum states},}\ }in\ \href@noop {} {\emph {\bibinfo
  {booktitle} {Optical Fiber Communication Conference and International
  Conference on Quantum Information}}}\ (\bibinfo  {publisher} {Optical Society
  of America},\ \bibinfo {year} {2001})\ p.\ \bibinfo {pages}
  {PB28}\BibitemShut {NoStop}%
\bibitem [{\citenamefont {McClean}\ \emph {et~al.}(2016)\citenamefont
  {McClean}, \citenamefont {Romero}, \citenamefont {Babbush},\ and\
  \citenamefont {Aspuru-Guzik}}]{mcclean_theory_2016}%
  \BibitemOpen
  \bibfield  {author} {\bibinfo {author} {\bibfnamefont {J.~R.}\ \bibnamefont
  {McClean}}, \bibinfo {author} {\bibfnamefont {J.}~\bibnamefont {Romero}},
  \bibinfo {author} {\bibfnamefont {R.}~\bibnamefont {Babbush}}, \ and\
  \bibinfo {author} {\bibfnamefont {A.}~\bibnamefont {Aspuru-Guzik}},\
  }\bibfield  {title} {\enquote {\bibinfo {title} {The theory of variational
  hybrid quantum-classical algorithms},}\ }\href@noop {} {\bibfield  {journal}
  {\bibinfo  {journal} {New J. Phys.}\ }\textbf {\bibinfo {volume} {18}},\
  \bibinfo {pages} {023023} (\bibinfo {year} {2016})}\BibitemShut {NoStop}%
\bibitem [{\citenamefont {Kandala}\ \emph {et~al.}(2017)\citenamefont
  {Kandala}, \citenamefont {Mezzacapo}, \citenamefont {Temme}, \citenamefont
  {Takita}, \citenamefont {Brink}, \citenamefont {Chow},\ and\ \citenamefont
  {Gambetta}}]{kandala_hardware-efficient_2017}%
  \BibitemOpen
  \bibfield  {author} {\bibinfo {author} {\bibfnamefont {A.}~\bibnamefont
  {Kandala}}, \bibinfo {author} {\bibfnamefont {A.}~\bibnamefont {Mezzacapo}},
  \bibinfo {author} {\bibfnamefont {K.}~\bibnamefont {Temme}}, \bibinfo
  {author} {\bibfnamefont {M.}~\bibnamefont {Takita}}, \bibinfo {author}
  {\bibfnamefont {M.}~\bibnamefont {Brink}}, \bibinfo {author} {\bibfnamefont
  {J.~M.}\ \bibnamefont {Chow}}, \ and\ \bibinfo {author} {\bibfnamefont
  {J.~M.}\ \bibnamefont {Gambetta}},\ }\bibfield  {title} {\enquote {\bibinfo
  {title} {Hardware-efficient variational quantum eigensolver for small
  molecules and quantum magnets},}\ }\href@noop {} {\bibfield  {journal}
  {\bibinfo  {journal} {Nature}\ }\textbf {\bibinfo {volume} {549}},\ \bibinfo
  {pages} {242} (\bibinfo {year} {2017})}\BibitemShut {NoStop}%
\bibitem [{\citenamefont {Moll}\ \emph {et~al.}(2018)\citenamefont {Moll},
  \citenamefont {Barkoutsos}, \citenamefont {Bishop}, \citenamefont {Chow},
  \citenamefont {Cross}, \citenamefont {Egger}, \citenamefont {Filipp},
  \citenamefont {Fuhrer}, \citenamefont {Gambetta}, \citenamefont {Ganzhorn},
  \citenamefont {Kandala}, \citenamefont {Mezzacapo}, \citenamefont
  {M{\"u}ller}, \citenamefont {Riess}, \citenamefont {Salis}, \citenamefont
  {Smolin}, \citenamefont {Tavernelli},\ and\ \citenamefont
  {Temme}}]{moll_quantum_2018}%
  \BibitemOpen
  \bibfield  {author} {\bibinfo {author} {\bibfnamefont {N.}~\bibnamefont
  {Moll}}, \bibinfo {author} {\bibfnamefont {P.}~\bibnamefont {Barkoutsos}},
  \bibinfo {author} {\bibfnamefont {L.~S.}\ \bibnamefont {Bishop}}, \bibinfo
  {author} {\bibfnamefont {J.~M.}\ \bibnamefont {Chow}}, \bibinfo {author}
  {\bibfnamefont {Andrew}\ \bibnamefont {Cross}}, \bibinfo {author}
  {\bibfnamefont {D.~J.}\ \bibnamefont {Egger}}, \bibinfo {author}
  {\bibfnamefont {S.}~\bibnamefont {Filipp}}, \bibinfo {author} {\bibfnamefont
  {A.}~\bibnamefont {Fuhrer}}, \bibinfo {author} {\bibfnamefont {J.~M.}\
  \bibnamefont {Gambetta}}, \bibinfo {author} {\bibfnamefont {M.}~\bibnamefont
  {Ganzhorn}}, \bibinfo {author} {\bibfnamefont {A.}~\bibnamefont {Kandala}},
  \bibinfo {author} {\bibfnamefont {A.}~\bibnamefont {Mezzacapo}}, \bibinfo
  {author} {\bibfnamefont {P.}~\bibnamefont {M{\"u}ller}}, \bibinfo {author}
  {\bibfnamefont {W.}~\bibnamefont {Riess}}, \bibinfo {author} {\bibfnamefont
  {G.}~\bibnamefont {Salis}}, \bibinfo {author} {\bibfnamefont
  {J.}~\bibnamefont {Smolin}}, \bibinfo {author} {\bibfnamefont
  {I.}~\bibnamefont {Tavernelli}}, \ and\ \bibinfo {author} {\bibfnamefont
  {K.}~\bibnamefont {Temme}},\ }\bibfield  {title} {\enquote {\bibinfo {title}
  {Quantum optimization using variational algorithms on near-term quantum
  devices},}\ }\href@noop {} {\bibfield  {journal} {\bibinfo  {journal}
  {Quantum Sci. Tech.}\ }\textbf {\bibinfo {volume} {3}},\ \bibinfo {pages}
  {030503} (\bibinfo {year} {2018})}\BibitemShut {NoStop}%
\bibitem [{\citenamefont {Kokail}\ \emph {et~al.}(2019)\citenamefont {Kokail},
  \citenamefont {Maier}, \citenamefont {van Bijnen}, \citenamefont {Brydges},
  \citenamefont {Joshi}, \citenamefont {Jurcevic}, \citenamefont {Muschik},
  \citenamefont {Silvi}, \citenamefont {Blatt}, \citenamefont {Roos},\ and\
  \citenamefont {Zoller}}]{kokail_self-verifying_2019}%
  \BibitemOpen
  \bibfield  {author} {\bibinfo {author} {\bibfnamefont {C.}~\bibnamefont
  {Kokail}}, \bibinfo {author} {\bibfnamefont {C.}~\bibnamefont {Maier}},
  \bibinfo {author} {\bibfnamefont {R.}~\bibnamefont {van Bijnen}}, \bibinfo
  {author} {\bibfnamefont {T.}~\bibnamefont {Brydges}}, \bibinfo {author}
  {\bibfnamefont {M.~K.}\ \bibnamefont {Joshi}}, \bibinfo {author}
  {\bibfnamefont {P.}~\bibnamefont {Jurcevic}}, \bibinfo {author}
  {\bibfnamefont {C.~A.}\ \bibnamefont {Muschik}}, \bibinfo {author}
  {\bibfnamefont {P.}~\bibnamefont {Silvi}}, \bibinfo {author} {\bibfnamefont
  {R.}~\bibnamefont {Blatt}}, \bibinfo {author} {\bibfnamefont {C.~F.}\
  \bibnamefont {Roos}}, \ and\ \bibinfo {author} {\bibfnamefont
  {P.}~\bibnamefont {Zoller}},\ }\bibfield  {title} {\enquote {\bibinfo {title}
  {Self-verifying variational quantum simulation of lattice models},}\
  }\href@noop {} {\bibfield  {journal} {\bibinfo  {journal} {Nature}\ }\textbf
  {\bibinfo {volume} {569}},\ \bibinfo {pages} {355} (\bibinfo {year}
  {2019})}\BibitemShut {NoStop}%
\bibitem [{\citenamefont {McClean}\ \emph {et~al.}(2018)\citenamefont
  {McClean}, \citenamefont {Boixo}, \citenamefont {Smelyanskiy}, \citenamefont
  {Babbush},\ and\ \citenamefont {Neven}}]{mcclean_barren_2018}%
  \BibitemOpen
  \bibfield  {author} {\bibinfo {author} {\bibfnamefont {J.~R.}\ \bibnamefont
  {McClean}}, \bibinfo {author} {\bibfnamefont {S.}~\bibnamefont {Boixo}},
  \bibinfo {author} {\bibfnamefont {V.~N.}\ \bibnamefont {Smelyanskiy}},
  \bibinfo {author} {\bibfnamefont {R.}~\bibnamefont {Babbush}}, \ and\
  \bibinfo {author} {\bibfnamefont {H.}~\bibnamefont {Neven}},\ }\bibfield
  {title} {\enquote {\bibinfo {title} {Barren plateaus in quantum neural
  network training landscapes},}\ }\href@noop {} {\bibfield  {journal}
  {\bibinfo  {journal} {Nature Communications}\ }\textbf {\bibinfo {volume}
  {9}},\ \bibinfo {pages} {4812} (\bibinfo {year} {2018})}\BibitemShut
  {NoStop}%
\bibitem [{\citenamefont {Grant}\ \emph {et~al.}(2019)\citenamefont {Grant},
  \citenamefont {Wossnig}, \citenamefont {Ostaszewski},\ and\ \citenamefont
  {Benedetti}}]{grant_initialization_2019}%
  \BibitemOpen
  \bibfield  {author} {\bibinfo {author} {\bibfnamefont {E.}~\bibnamefont
  {Grant}}, \bibinfo {author} {\bibfnamefont {L.}~\bibnamefont {Wossnig}},
  \bibinfo {author} {\bibfnamefont {M.}~\bibnamefont {Ostaszewski}}, \ and\
  \bibinfo {author} {\bibfnamefont {M.}~\bibnamefont {Benedetti}},\ }\bibfield
  {title} {\enquote {\bibinfo {title} {An initialization strategy for
  addressing barren plateaus in parametrized quantum circuits},}\ }\href@noop
  {} {\bibfield  {journal} {\bibinfo  {journal} {arXiv:1903.05076 [quant-ph]}\
  } (\bibinfo {year} {2019})}\BibitemShut {NoStop}%
\end{thebibliography}
\end{document}